\documentclass[
reprint,
superscriptaddress,
secnumarabic, graphics,floatfix,nofootinbib,
tightenlines,nobibnotes,
amsmath,amssymb,
aps,
prx,
]{revtex4-1}

\usepackage[utf8]{inputenc}
\usepackage[T1]{fontenc} 
\usepackage[english]{babel} 

\usepackage{cancel,xcolor}
\usepackage{dsfont}
\usepackage{appendix}
\usepackage[normalem]{ulem}
\usepackage{enumitem}
\usepackage{blindtext}
\usepackage{comment}

\usepackage{braket}
\usepackage{amsmath}
\usepackage{amssymb}
\usepackage{amsfonts}
\usepackage{amsthm}
\usepackage{graphicx}
\usepackage[scale=0.8]{geometry}
\usepackage{float}

\definecolor{dark-green}{RGB}{0, 128, 0}
 \newcommand{\piR}[1]{\textcolor{dark-green}{#1}}

\newcommand{\ii}{\mathrm{i}}

\newcommand{\chern}{\text{Ch}}
\newcommand{\ind}{\text{Ind}}

\newcommand{\sign}{\text{sgn}}

\usepackage{fourier}

\begin{document}
\title{ Accessing Chern Numbers From Bloch Eigenstates Singularities}
\author{Rajesh Asapanna}
\author{Rabih El Sokhen}
\affiliation{Univ. Lille, CNRS, UMR 8523 -- PhLAM -- Physique des Lasers Atomes et Mol\'ecules, F-59000 Lille, France}

\author{Paolo~Aceto}
\author{Patrick Popescu-Pampu}
\affiliation{Univ. Lille, CNRS, UMR 8524 - Laboratoire Paul Painlevé, F-59000 Lille, France}

\author{Martin~Guillot}
\author{Jacqueline~Bloch}
\author{Sylvain~Ravets}
\affiliation{Université Paris-Saclay, CNRS, Centre de Nanosciences et de Nanotechnologies, 91120 Palaiseau, France}

\author{Clément~Hainaut}
\author{Alberto~Amo}
\email{alberto.amo-garcia@univ-lille.fr}
\affiliation{Univ. Lille, CNRS, UMR 8523 -- PhLAM -- Physique des Lasers Atomes et Mol\'ecules, F-59000 Lille, France}

\author{Pierre~Delplace}
\email{pierre.delplace@ens-lyon.fr}
\affiliation{CNRS, ENS de Lyon, LPENSL, UMR5672, 69342, Lyon cedex 07, France}

\begin{abstract}
We propose two methods to extract the Chern numbers of the Bloch bands of a two dimensional insulator, and apply them to a photonic lattice experiment. These two methods require the knowledge of the complex-valued components of the Bloch eigenvectors, or their ratios. Unlike other tomography methods relying on the approximate reconstruction of the Berry curvature and its integration over the Brillouin zone, our methods boil down to the observation of phase vorticities in the eigensates structure. Those measurements, robust to experimental noise, yield exactly quantized values. One of the two methods exploits the non-normalization of the eigenmodes, making it particularly  relevant for classical wave systems.
\end{abstract}

\maketitle
\section{Introduction} Chern numbers are probably the most emblematic topological invariants used in condensed matter physics. Introduced in mathematics in the mid-XXth century by Shiing-Shen Chern to classify abstract complex vector bundles \cite{Chern46}, they were much later recognized by Thouless, Kohmoto, den Nijs and Nightingale to explain the surprizingly robust quantization of the freshly discovered plateaus of the quantum Hall effect~\cite{Thouless1982}. In this topological interpretation of the quantized Hall effect, the transverse conductivity $\sigma_H$ is found to be proportional to the Chern number $C$ of a $U(1)$-bundle associated to the Bloch states over the Brillouin zone, in units of the quantum of conductance, i.e. $\sigma_H=(e^2/h) C$. Measuring a response function thus gives access to  information about the topological properties of the ground state. Such a direct relation between a physical observable and a topological invariant is one of the quests of topological physics. The prediction in Weyl semimetals of the quantization of a circular galvanic effect in units of $e^3/h^2$, also proportional to a Chern number, is a remarkable success along this line \cite{Juan2017}. 

In recent years, various experimental platforms have emerged to investigate the physics of wavebands, characterized by Chern numbers. Classical waves are particularly well suited for this endeavor, and topological metamaterials have been realized in photonics, acoustics, mechanics and hydrodynamics. In such classical wave systems, the existence of topologically quantized response functions related to those topological indices remains however unlikely. In order to unveil their topological properties, most studies then focused on the existence of robust unidirectional edge states through the bulk-boundary correspondence, thus providing an indirect signature of the bulk topology~\cite{Hafezi2013, Rechtsman2013b, susstrunk_observation_2015,Delplace2017, xue_topological_2022}. Few other studies investigated the Berry curvature of the bulk bands, from either anomalous wave-packet velocities~\cite{Aidelsburger2015, li2016, Wimmer2017, Wintersperger2020, Gianfrate2020, Chen2023, chenier_quantized_2024}, adiabatic transport~\cite{chen_direct_2025} or tomography measurements~\cite{Flaschner2016, Flaschner2018, Tarnowski2019, Yu2020, Yi2023, el_sokhen_edge-dependent_2024,guillot_sublattice_2025}. In all those cases, an approximate integer value of the expected Chern number can then be inferred after integration over the Brillouin zone. 

Here, we follow a different strategy to directly access the integer-valued Chern numbers of the bands. We search, in the eigenstate structure, manifestations of the original mathematical meaning  of the first Chern numbers as topological obstructions of a complex vector bundle. Such signatures are found in eigenstates singularities,  where we take advantage of a key difference between classical and quantum Bloch wavefunctions, which is that the former  need not be normalized. Interestingly, the singularities displayed by non-normalized wavefunctions are countable objects, thus providing direct access to quantized quantities. This contrasts with other approaches that rely on numerically integrating the Berry curvature obtained from (normalized) Bloch wavefunctions, and that yield an approximate value of the quantized first Chern number. We emphasize that our approaches are general and can be applied to any classical wave system where eigensate tomography is possible, and we demonstrate their practicability in a photonic lattice experiment.

We thus propose two methods to extract the Chern numbers without resorting to the Berry curvature. These two methods are based on different mathematical objects. The first one is based on the identification of the zeros of a continuous section of a complex line bundle; the second one on the degree of a map between the Brillouin zone and the projective space. Those two methods require the complete knowledge of the eigenvectors and are both accessible in our experiment based on a time multiplexed lattice for light pulses in which we implement various two-dimensional two-band models. One of the crucial advantages of the two methods compared to measurements of the Berry curvature or the anomalous transport over the entire Brillouin zone is that it is based on the identification of zeros of the different components of the eigenvectors (or their ratio) in the Brillouin zone and on the associated phase circulation around those points which is a quantity very robust  to experimental noise.

The paper is organized as follows: First, a theoretical section  details our two methods that we dub the \textit{0-section method} and the \textit{tautological bundle method}. The Haldane model for a Chern insulator is revisited to illustrate those methods. The second section describes our experimental setups and the implemented band models. The Chern numbers of the Bloch bands are then extracted experimentally by following each of the two methods, and compared to a measurement of the Berry curvature.



\section{Chern numbers from Bloch bundles}
\subsection{0-section method: theory} 

The first method we propose follows from a longstanding observation that Bloch eigenstates can be mathematically interpreted as  sections of a complex bundle. 
In particular, it is well known in mathematics that when dealing with complex line bundles over surfaces (such as the Brillouin zone torus in two dimensions), the first Chern number counts the number of zeros -- with signs -- of generic sections \cite{Chern46}. In the context of the quantum Hall effect, this abstract statement was interpreted by Kohmoto as the fact that the complex wavefunction must vanish in the Brillouin zone, and those vanishings come with a phase singularity with integer vorticity; the Chern number is then the sum of those vorticities~\cite{Kohmoto1985}.
To our knowledge, this interpretation of the Chern numbers never gave rise to a protocol to measure them. Indeed, a first difficulty we may see is that quantum wavefunctions being normalized for each quasi-momentum, they actually cannot vanish in the Brillouin zone. This obstacle is naturally avoided if we consider instead classical waves, since their vector states are not subject to physical normalization constraints. On the contrary, the norm of classical vector states is more likely related to the energy of the wave excitation. For that reason, we shall focus on non-normalized eigenstates. Still, a second difficulty appears, which is that a zero of a multi-component complex-valued vector state implies the vanishing of several complex functions (each component of the vector), which may come with different vorticies of the phase. It is then not clear which vorticity should be taken into account. In other words, how to attribute a sign to a zero of a multi-component vector state?

To answer this question and clarify a protocol to extract the first Chern number from the zeros of the Bloch wavefunction in the Brillouin zone, let us first recall the construction of complex vector bundles in the context of Bloch waves.

Consider a Bloch eigenstate $\psi_n(k)$ of a two-dimensional crystal, where $n="1, \dots ,N"$ is the band index and $k$ is the quasi-momentum that lives in the Brillouin zone (BZ), and let us assume no degeneracy of the bands.
The use of vector bundle theory in this context follows from two observations: (1) BZ is an oriented two-dimensional torus and, thus, a closed (compact) manifold, such that the complex-valued vector states $\psi_n(k)$ can be seen as maps from this closed manifold to $\mathbb{C}^N$; (2) for any $k \in $ BZ, the choice of $\psi_n(k)$ is not unique, as any multiplication by a non-zero complex number $\lambda$ yields a physically valid eigenvector $\psi'_n(k)=\lambda \psi_n(k)$. This is nothing but saying that, for a given $k$, any fixed Bloch eigenvector $\psi_n(k)$ lives in a complex eigenspace of dimension $1$ "parametrized" by $\lambda$, that we denote by $L_n(k)$. 
Seen as a function of $k\in$ BZ, $\psi_n(k)$ lives in a continuous collection of parametrized eigenspaces $\mathbb{L}_n$. In the language of vector bundles, this translates as the existence of a projection $\Pi_{BZ}^n:\mathbb{L}_n\rightarrow$ BZ that associates a common quasi-momentum $k \in $ BZ to all the Bloch eigenvector $\psi'_n(k)=\lambda \psi_n(k)$. This projection defines a Bloch bundle for each of the N eigen-subspaces of the Bloch Hamiltonian. It is a complex line bundle that formally reads 
\begin{align}
    \mathbb{L}_n = \{(k,\psi_n)\in \text{BZ} \times \mathbb{C}^N|\psi_n \in (\Pi^n_{BZ})^{-1}(k) = L_n(k)  \} \ .
    \label{eq:line_bundle}
\end{align}
A section of the line bundle $\mathbb{L}_n$ is a map from BZ to $\mathbb{L}_n$ sending each point $k$ of  BZ to a point of the corresponding line $L_n(k)$. Of importance are continuous sections: continuous choices of vectors $\psi_n(k)$ in each $L_n(k)$ for every $k$. This gives the Bloch eigenvectors a geometrical meaning in the language of line bundles. Importantly, it is well known that the first Chern number counts the zeros -- with signs -- of continuous sections of line bundles (see e.g. p 467 of \cite{Frankel}). Thus, the existence of a non-zero Chern number $C_n\equiv \chern(\mathbb{L}_n)$ of a Bloch bundle $\mathbb{L}_n$ should be accessible from the zeros of a continuous Bloch wavefunction $\psi_n(k)$. Let us assume that we have a finite number of them, and let us denote by $\bar{k}_j$ the points in BZ where an eigenvector vanishes, $\psi_n(\bar{k}_j)=0$, that is, where all components of $\psi_n(k)$ are simultanously equal to zero. Recall that this is possible in the problem we are treating because we are not assuming that $\psi_n(k)$ is normalised. We then need to attribute a sign to these zeros to compute the Chern number. Let us see how.


\begin{figure*}[t!]
\centering
\includegraphics[width=1\textwidth]{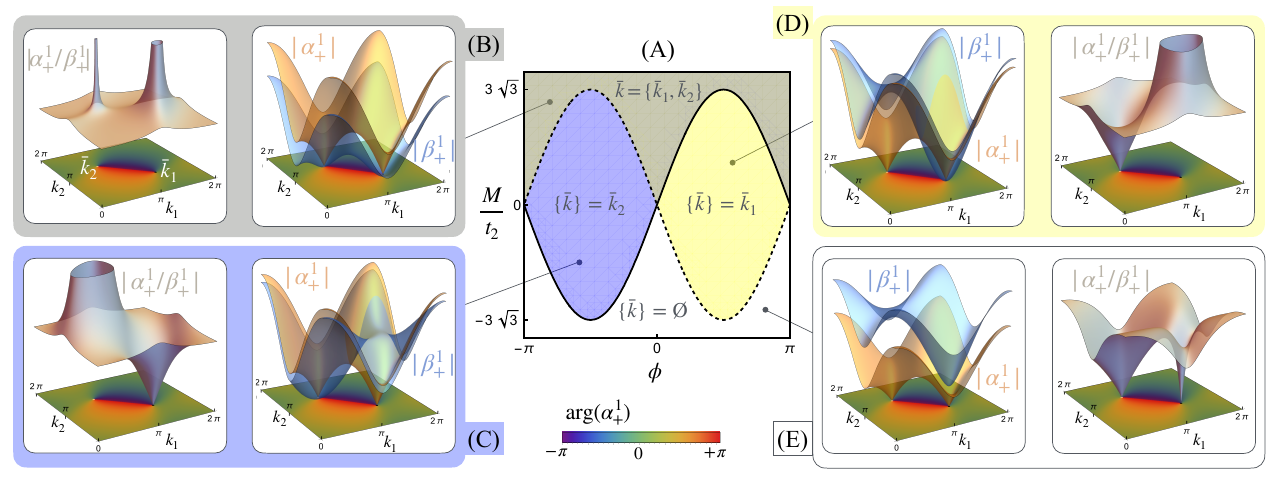}
\caption{(A) Phase diagram of the Haldane model obtained from the two different methods. Each coloured domain is characterized by a number of zeros $\bar{k}_j$ of a section, here $\psi_+^1$, or equivalently by the number of zeros $\bar{q}_j$ of the map $\mu_+^1$ (not shown). The curved surfaces in panels (B) to (E) show the value of the modulus of the components $\alpha_+^1$ (orange band) and $\beta_+^1$ (light blue band) of $\psi_+^1$, and the modulus of the map $\mu_+^1=\alpha_+^1/\beta_+^1$ (grey-reddish surface). The colour map displays the argument of respectively $\alpha_+^1$ and $\alpha_+^1/\beta_+^1$ as a function of $k_1$ and $k_2$ in BZ. The Chern number $C_+$ can then be inferred both from the winding of $\arg \alpha_+^1$ around the points where both $\alpha_+^1$ and $\beta_+^1$ vanish ($0$-section method), and from the winding of $\arg \mu_+^1$ around the zeros of $\mu_+^1$ (tautological bundle method). }
\label{fig:Haldane_section}
\end{figure*} 

For a  continuously differentiable real function $f(k)$, the sign of an isolated root $\bar{k}$ is given by the sign of its derivative $(df/dx)(\bar{k})$ at this point. For a complex function of two variables $\sigma(k_1,k_2)=X(k_1,k_2)+iY(k_1,k_2)$, this generalizes as the index of this function, defined by
\begin{equation}
\ind_{\bar{k}_j} \sigma=\sign \left[\det 
\begin{pmatrix}
    \frac{\partial X}{\partial k_1} & \frac{\partial X}{\partial k_2}  \\
    \frac{\partial Y}{\partial k_1} & \frac{\partial Y}{\partial k_2} 
\end{pmatrix}
\right] 
\label{eq:ind_jacobian}
\end{equation}
evaluated at a (non-degenerate) zero $\bar{k}_j$ of $\sigma$. It does not make sense to use directly this formula with $\psi_n(k)$  because $\psi_n(k)$ is not a scalar complex function, but a vector-valued complex function. However, this difficulty can be overcome because we deal with line bundles, meaning that each vector in each one-dimensional vector space $L_n(k)$ can be identified with a complex number. The procedure that locally reduces $\psi_n(k)$ to a scalar complex function $\sigma_n(k)$ in a vicinity of a zero of $\psi_n(k)$ is called \textit{trivialization} \cite{Husemoller}. This process is not unique. A possible choice of trivialization consists in keeping only the component of $\psi_n(k)$ that vanishes the slowest when approaching $\bar{k}_j$. The component that is kept might be different for each $\bar{k}_j$. %
Following this procedure, that we detail below,  the Chern number of the n-th band is given by 
\begin{align}
C_n = \sum_j \text{ind}_{\bar{k}_j}\sigma_{n}.
\label{eq:chern_ind}
\end{align}
Note that the ordering of the coordinates $(k_1, k_2)$ fixes the orientation of the BZ and thus the sign of $C_n$. Let us also mention that the indices are local quantities well defined close to each point $\bar{k}_j$: the component of the vector $\psi_n(k)$ that vanishes the slowest (trivilization) might be different at different $\bar{k}_j$, each of them sets the index at each point $\bar{k}_j$, and their sum (the Chern number) is independent of the choice of trivialization.

\begin{figure}[t!]
\includegraphics[width=8.6cm]{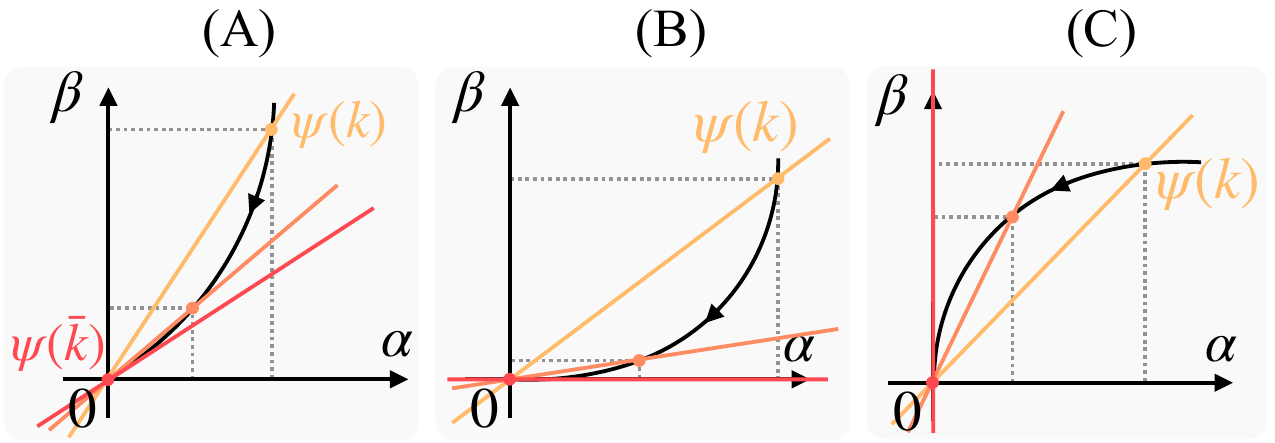}
\caption{\label{fig_trivialisation} Trajectories (oriented black lines) of a vector state $\psi(k)$ in $\mathbb{C}^2$ when $k$ approaches a zero  of $\psi(k)$ at $k=\bar{k}$. The different straight lines represent the associated 1-dimensional complex vector space $L(k)$ spanned by $\psi(k)$, for different $k$ along the trajectory. When approaching the zero, $L(k\rightarrow \bar{k})$ becomes tangent to the trajectory of $\psi$ at $0$ (red line), and may coincide with one coordinate axis (B,C). In that case, a local trivialization is obtained by projecting $\psi$ onto that axis, as a projection onto the other (perpendicular) axis vanishes. (A) generic case: a projection on either coordinate axis provides a local trivialization of the section.  } 
\end{figure}

From now on, let us focus on two-band models for simplicity, and let us write $\psi_\pm(k)=(\alpha_\pm(k),\beta_\pm(k))^T$ where $T$ is the transpose and $\alpha_\pm(k)$ and $\beta_\pm(k)$ are the complex coordinates of the eigenvectors. To identify a local section $\sigma_\pm(k)$, we could project $\psi_n(k) $ onto a given complex line in $\mathbb{C}^2$. Let us choose the axes of coordinates $\alpha$ and $\beta$ as candidates for this projection. We need to check that when $k$ approaches $\bar{k}_j$, that is when $\psi(k)\rightarrow 0$, this projection is well defined, which amounts to consider $\lim_{k \to \bar{k}_j} \beta(k)/\alpha(k)$. In general, this limit yields a finite complex number, so that either $\alpha_\pm(k)$ or $\beta_\pm(k)$ can be taken as a local section in the vicinity of $\bar{k}_j$ (Fig. \ref{fig_trivialisation} (A)). But it may happen that this limit is $0$ or diverges, indicating that one of the two axes cannot be used to project. More precisely, if this limit is $0$ (Fig. \ref{fig_trivialisation} (B)), then the complex line whose $\psi$ belongs to reaches the $\alpha$ coordinates axis, implying that $\alpha_\pm(k)$ is a suitable local section, while the projection onto $\beta$ vanishes. If, on the contrary, this limit is $\infty$ (Fig. \ref{fig_trivialisation} (C)), then the complex line whose $\psi$ belongs to coincides with the $\beta$ coordinates axis, implying that $\beta\pm(k)$ is a suitable local section in that limit, while $\alpha_\pm(k)$ is not.   
In each case, the index can  be computed from the sign of the determinant of the Jacobian matrix as given by  Eq.~\eqref{eq:ind_jacobian}. 
We end up with the rule
\begin{align}
   \ind_{\bar{k}_j}\sigma_\pm = \left\{ \begin{array}{ll}
     W_{\bar{k}_j}[\alpha_\pm]=W_{\bar{k}_j}[\beta_\pm] & \text{if}\quad \beta_\pm/\alpha_\pm \to    \lambda \in \mathbb{C} \setminus \{0\} \\
   W_{\bar{k}_j}[\alpha_\pm]  &  \text{if}\quad \beta_\pm/\alpha_\pm \to 0 \\
   W_{\bar{k}_j}[\beta_\pm]  & \text{if}\quad \beta_\pm/\alpha_\pm \to \infty
   \end{array}
    \right.
\label{eq:ind_wind}
\end{align}
where $\to$ denotes the limit when $k\to \bar{k}_j$.
Equations \eqref{eq:chern_ind} and \eqref{eq:ind_wind} constitute a first procedure to evaluate  the Chern numbers from the wavefunction components. This procedure generalizes straightforwardly to eigenstates $\psi_n(k)=(\alpha_1, \dots , \alpha_N)^T\in \mathbb{C}^N$ of an $N$-band insulator, where the eigenstate's component $\alpha_j(k)$ can be chosen as a local trivialization if $\alpha_j(k)/\alpha_i(k) \rightarrow \{\infty$ or $\lambda \in \mathbb{C}\}$ when $\alpha_i$ is any of the $N-1$ other components.  

In practice, as we illustrate below, the expression of $\psi_n(k)$ is gauge dependent, which makes the number of zeros also gauge dependent, but the sum \eqref{eq:chern_ind} of their indices  is invariant. Furthermore, the index is equivalently given by the winding number $W_j[\sigma]=(2\pi i)^{-1}\oint dk (\sigma^*\partial_k \sigma)|/\sigma|^2$ of  $\arg \sigma_\pm(k)$ along a closed loop around the zero $\bar{k}_j$ of $\psi_\pm(k)$ in the Brillouin zone. This expression will be of practical interest in the following because it can be measured experimentally via the phase of the trivialization component of the wavefunction around that point.

It is worth stressing that that the knowledge of the phase winding around the phase singularities in the Brillouin zone is not sufficient in itself to infer the Chern number. Actually, in crystals, the phase singularities of the complex components of the Bloch eigenvectors have to appear in pairs of opposite winding numbers, so that their sum is always zero. This is a direct consequence of the fact that the Brillouin torus is a compact manifold. Indeed, a component of a Bloch eigenvector being a map $z:\mathbb{T}^2\rightarrow \mathbb{C}$, it follows that $z(\mathbb{T}^2)$ is bounded and closed. 
The degree of this map is therefore zero since there are elements of the target space $\mathbb{C}$ which do not have preimages in BZ. Since the degree is the sum of the winding numbers around the points $\bar{k}_j$ where $z$ vanishes, it follows that the sum of these winding numbers vanishes. The 0-section method thus provides us with a way to select the phase singularities that must be kept in the counting of the indices appearing in relation \eqref{eq:chern_ind} (that is where the state vector vanishes), from those that must be disregarded. 

\subsection{Application to the Haldane model}

Before discussing a second method for evaluating the Chern number, let us  illustrate the \textit{0-section method} with the seminal Haldane model of a Chern insulator \cite{Haldane88}, which is a two-band model Hamiltonian $H(k)=\mathbf{h}(k)\cdot \boldsymbol{\sigma}$ with $k=(k_x,k_y)$,  $\boldsymbol{\sigma}=(\sigma_x,\sigma_y,\sigma_z)$ the Pauli matrices and  $h_x(k)=1+\cos k_x+\cos k_y$, $h_y(k)=-\sin k_x-\sin k_y$, $h_z(k)=M+2t_2\sin\phi[\sin(k_x-k_y)-\sin(k_x+k_y)+\sin k_y]$. The parameters $M/t_2$ and $\phi$ control the well-known topological phase diagram that we aim at reproducing by using \eqref{eq:chern_ind} and \eqref{eq:ind_wind}. The eigenspaces $V_\pm(k)$, associated to the eigenvalues $E_\pm(k) = \pm|\mathbf{h}(k)|\equiv\pm E$
 are given by the system of equations
\begin{eqnarray}
\label{eq:gauge1}
&(h_z-E_\pm)\alpha_\pm + (h_x-\ii h_y) \beta_\pm \ =0 & \quad \text{(gauge 1)} \\
&(h_x+\ii h_y) \alpha_\pm + (-h_z -E_\pm) \beta_\pm =0 & \quad \text{(gauge 2)} \ .
\label{eq:gauge2}
\end{eqnarray} 
Each of these equations yields straightforwardly the two eigenvectors $\psi_\pm$ (up to a global multiplication factor), but in different \textit{gauges}
\begin{align}
\label{eq:psi_chart}
\psi_\pm^1 =
\begin{pmatrix} h_x-\ii h_y \\ E_\pm-h_z \end{pmatrix}
\qquad
\psi_\pm^2 = \begin{pmatrix} E_\pm+h_z \\ h_x+\ii h_y \end{pmatrix},
\end{align}
where the upper index labels the gauge.
As motivated in the introduction, we consider non-normalized eigenvectors as we should naturally do with classical waves. The normalization of the eigenvectors would prevent their vanishing, making the \textit{0-section method} irrelevant.
The reason why we refer to the different expressions~\eqref{eq:psi_chart} as two different gauge choices is that they are related to each other by multiplication with a non-zero complex number: $\psi_\pm^2= \lambda_\pm \psi_\pm^1 $ with $\lambda_\pm=(E_\pm-h_z)/(h_x-\ii h_y)$ when $E_\pm\neq h_z$. This is analogous to the standard gauge transform with normalized eigenstates that would consist in a multiplication by $\lambda$ with $|\lambda| =1$). In the language of line bundles, the two gauges correspond to two different choices of section of the same line bundle \eqref{eq:line_bundle}.

Let us now compute the Chern number $C_+ = \chern(\mathbb{L}_+)$ from the parametrized eigenvector $\psi_+^1(k)$ in gauge 1. The vanishing condition $\psi_+^1(\bar{k}_j)=0$ translates into two equations: $\alpha_+^1(\bar{k}_j)=0$, which gives the possible locations $\bar{k}_1=(4\pi/3,2\pi/3)$ and $\bar{k}_2=(2\pi/3,4\pi/3)$ in the Brillouin zone, and $\beta_+^1(k_j)=0$ that imposes the constraint $h_z(\bar{k}_j)>0$. Substituting the values of $\bar{k}_1$ and $\bar{k}_2$ yields
\begin{align}
\psi^1_+(\bar{k}_1)=0\quad  &\text{for}\hspace{0.2cm}   M> -3\sqrt{3} \sin\phi    \label{eq:9}\\
\psi^1_+(\bar{k}_2)=0\quad  & \text{for}\hspace{0.2cm}    M> 3\sqrt{3} \sin\phi \ .
\label{eq:10}
\end{align}
Thus, depending on the values of $M$ and $\phi$, $\psi_+^1(k)$ has either none, one or two zeros $\bar{k}_j$, leading to the four domains shown in Fig. \ref{fig:Haldane_section} (A). 
In the white domain (panel (E)), there is no zero of $\psi_+^1$, and thus $C_+=0$. 
In the blue and yellow domains there is only one zero of $\psi^1_+(\bar{k})$, while in the grey domain there are two.
To extract the Chern number in those domains we use the local trivialization of  $\psi_+^1$ described above.
We find that $|\beta_+^1/\alpha_+^1| \rightarrow 0$ when $k$ approaches $\bar{k}_j$ in the three domains, which can be seen in Figs.~\ref{fig:Haldane_section} (C,D,E) at those points in the inverse condition ($|\alpha_+^1/\beta_+^1| \rightarrow \infty$. One can thus take $\alpha_+^1(k)$ as a local section in the vicinity of each $\bar{k}_j$, so that $C_+$ is finally given by the sum of the windings of $\arg(\alpha_+^1)$ around each $\bar{k}_j$. The argument $\arg(\alpha_+^1)$ is shown in color-code projected onto the BZ in panels (B-E). In the blue and yellow domains, the section vanishes once, respectively at $\bar{k}_2$ and $\bar{k}_1$, leading to $C_+=W_{\bar{k}_2}[\alpha_+^1]=+1$ in the blue domain and $C_+=W_{\bar{k}_1}[\alpha_+^1]=-1$ in the yellow domain. Finally, in the brown domain (panel B), the two winding numbers $W_{\bar{k}_1}[\alpha_+^1]=-1$ and $W_{\bar{k}_2}[\alpha_+^1]=+1$ compensate each other leading to $C_+=0$. This reproduces the celebrated topological phase diagram of the Haldane model.

\subsection{Tautological bundle method: theory} 

We now discuss a second method to evaluate the Chern numbers from the Bloch eigenvectors components. This approach relies on the fact (which is a consequence of Theorem 14.6 of \cite{Milnor}) that any complex line bundle on a surface -- such  as  Bloch  bundles with non-degenerate bands --  is the pullback of a fixed line bundle, called the \textit{tautological} bundle $\mathbb{L}_0$, so that $\chern(\mathbb{L}_n)=\chern(\mathbb{L}_0) \deg{\mu_n}$
where $\mu_n$ is a map from the base space BZ of our Bloch bundle to the base space of the tautological bundle. Since the tautological bundle is known to have $\chern(\mathbb{L}_0)=-1$ \cite{Patrick}, we are left with
\begin{equation}
    C_\pm = - \deg{\mu_\pm} 
    \label{eq:chern_deg}
\end{equation}
for our two-band Hamiltonian. This degree formula should not be mistaken with another currently used degree formula to compute the Chern numbers of a two-band Hamiltonian where the map $\mathbf{h} : \text{BZ} \rightarrow \mathbb{R}^3/\{0\}$, which is considered in that case, encodes the Bloch Hamiltonian as $H(k)=\mathbf{h}(k)\cdot \boldsymbol{\sigma}$ \cite{Sticlet12, FruchartCRAS}. 

To identify the maps $\mu_\pm$, let us first recall what the tautological bundle is and explain how it appears in our setting. 
For this, let us go back to Eqs. \eqref{eq:gauge1} and \eqref{eq:gauge2}, and consider them now as equations defining lines in $\mathbb{C}^2$ passing through the origin. Any non-zero vector state  $\psi\in\mathbb{C}^2$ belongs to such a line.  The space of all those lines is called the complex projective line $\mathbb{C}\mathbb{P}^1$ and constitutes the base space of the tautological bundle $\mathbb{L}_0$. A point $p \in \mathbb{C}\mathbb{P}^1$ thus represents a line $L_0(p)\in \mathbb{C}^2$ through the projection $\Pi_{\mathbb{C}\mathbb{P}^1}:\mathbb{L}_0\rightarrow \mathbb{C}\mathbb{P}^1$ and the tautological bundle reads
\begin{align}
\mathbb{L}_0 = \{ (p,\psi) \in \mathbb{C}\mathbb{P}^1\times \mathbb{C}^2 | \psi\in \Pi_{\mathbb{C}\mathbb{P}^1}^{-1}(p)=L_0(p) \} \ .
\end{align}
Note that  \textit{two} equations, \eqref{eq:gauge1} and \eqref{eq:gauge2}, are necessary to describe all the lines in $\mathbb{C}^2$ passing through the origin. This is reminiscent of the topology of $\mathbb{C}\mathbb{P}^1$ which is the Riemann sphere $S^2$ and thus requires two charts to be defined with coordinates ($\alpha,\beta$). Those two charts correspond here to our two gauge choices:  Eq. \eqref{eq:gauge1} describes all such lines except that given by $\beta=0$ (when $E_\pm=h_z$), while Eq. \eqref{eq:gauge2} describes all those lines except that given by $\alpha=0$ (when $E_\pm=-h_z$).

\begin{figure*}[t!]
\includegraphics[width=\textwidth]{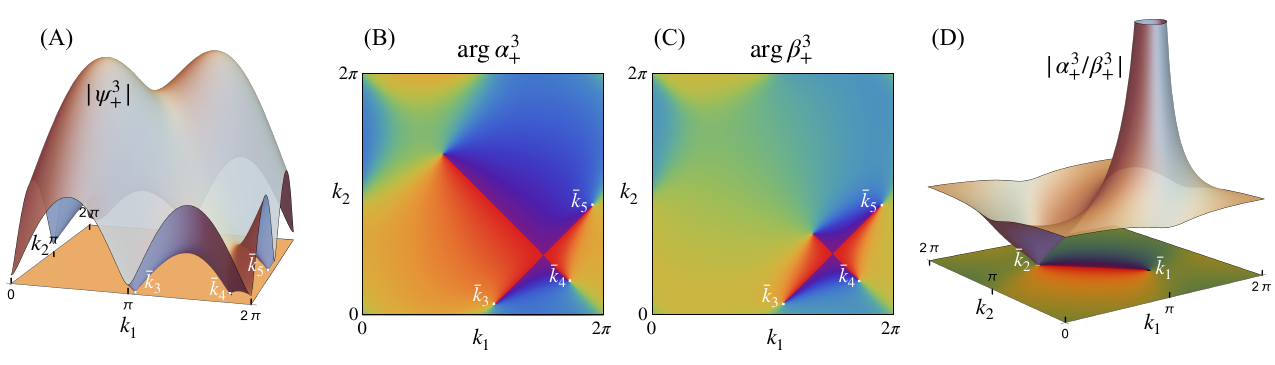}
\caption{\label{fig_jauge3} Zeros-section and tautological bundle methods applied to determine the Chern number in gauge 3 for the Haldane model (yellow domain of Fig. \ref{fig:Haldane_section}, $C_+=-1$). (A) The norm of the eigenvector $\psi_+^3$ in gauge 3 vanishes at three different points ($\bar{k}_3,\bar{k}_4,\bar{k}_5$) in BZ. (D) At those points, the limit $\alpha_+^3/\beta_+^3$ is finite, so any of $\arg \alpha_+^3$ (B) or $\arg \beta_+^3$ (C) can be chosen as a local trivialized section to determine the indices at those points. The color code for $\arg \alpha_+^3$ and $\arg \beta_+^3$ is the same as in Fig. \ref{fig:Haldane_section}}
\end{figure*}

Finally, concrete expressions of the  maps $\mu_\pm :\text{BZ} \rightarrow \mathbb{C}\mathbb{P}^1$ that associate to each $k$ of the BZ, a complex line in $\mathbb{C}^2$, can be identified over patches of the Brillouin zone, as $\mu^1_\pm=\alpha_\pm^1/\beta_\pm^1$ in gauge 1, and $\mu^2_\pm=\beta_\pm^2/\alpha_\pm^2$ in gauge 2. In practice, the degree of $\mu_\pm$ can be evaluated from the winding number $W_{\bar{q}_j}[\mu_\pm]$ of $\arg \mu_\pm$ around the zeroes $\overline{q}_i$ of $\mu_{\pm}$ in BZ (which are  \textit{a priori} different from the zeros $\bar{k}_j$ of a section), so that we end up with our second rule
\begin{equation}
    C_\pm = - \sum_j W_{\bar{q}_j}[\mu_\pm] \ .
    \label{eq:chern_tauto}
\end{equation}

We  illustrate this method with $\mu_+^1(k)$ that we plot in panels (B) to (E) of Fig.~\ref{fig:Haldane_section} for the Haldane model, for which $W_{\bar{q}_j}[\mu_+^1]$ can be read out directly in each domain from $\arg \mu_+^1(k)$ around the zeros of $\mu_+^1(k)$. 
The formula \eqref{eq:chern_tauto} consistently provides the same result for the Chern number as with the 0-section method. Note that the domains where $C_+=0$ can actually be inferred directly from Eq. \eqref{eq:chern_deg}. Indeed, panels (B) and (E) show that the map $\mu_+^1$ does not cover the Riemann sphere in those cases: $\mu_+^1=0$ is never reached in the brown domain (panel (B)), and $\mu_+^1=\infty$ is never reached in the white domain (panel (E)). Therefore, their degree is zero.

\subsection{Another gauge choice}

It is worth stressing that the \textit{0-section method} described above is gauge dependent, in the sense that the number and location of the zeros of a section $\psi^j_{\pm}$, depend on the gauge $j$. We chose gauge $j=$1 (i.e., Eq.~\eqref{eq:gauge1}) as an illustration, but the same analysis can be done in gauge 2 starting with Eq.~\eqref{eq:gauge2}. In that case, the position of the zeros and the winding of the arguments are just switched, yielding the same Chern numbers.

It may seem surprising that the zeros of interest always appear at the location of the gap closing points. This may seem related to the points of highest Berry curvature, which are those where the gap is the smallest, namely around the same points.  However, this is just a coincidence for the phase singularities that we are considering, precisely because the expression of the eigenvector components is gauge dependent. 

In this respect, it is instructive to consider a different gauge choice for which the section displays 3 zeros in the Chern phase, none of which coincides with the position of a gap closing  point, nor with the position of the zeros of $\mu$. Different yet equivalent continuous sections $\psi^j$ in a different gauge can be obtained by multiplying $\psi^1$ (or $\psi^2$) with any \textit{continuous} function $\lambda(k)$. For instance, multiplying $\psi^1$ by $\lambda(k)=e^{-i \arg{\alpha^1(k)}}$ is \textit{not} a valid gauge transformation as $\lambda(k)$ is not continuous at $\bar{k}_1$ and $\bar{k}_2$, that is where $\alpha^1(k)$ vanishes with a non-zero index. 

To illustrate a situation with 3 zeros in the Chern phase, we use a third gauge choice by summing Eqs~\eqref{eq:gauge1} and \eqref{eq:gauge2} so that the eigenvectors read as a linear combination of the eigenvectors in gauges 1 and 2 as
\begin{align}
\psi_\pm^3  
=\begin{pmatrix}h_z+ E_\pm - (h_x-\ii h_y) \\ h_z-E_\pm + h_x+ \ii h_y \end{pmatrix}
=\psi_\pm^2 - \psi_\pm^1 \, .
\end{align}
We can then apply the two methods above to extract the Chern number in this gauge. We show in Fig.~\ref{fig_jauge3} the relevant quantities to apply those methods to the upper band of the Haldane model in the range of parameters where $C_+=-1$. Panel A shows the norm $|\psi_+^3(k)|$, that reveals 3 zeros located at $\bar{k}_3$, $\bar{k}_4$ and $\bar{k}_5$, different from the previous $\bar{k}_1$ and $\bar{k}_2$ in gauges 1 and 2. Each of these zeros comes with a winding of the argument of each component ($\beta_+^3(k)$ is a complex-valued function in gauge 3). Then, panel D shows that the ratio $\alpha_+^3/\beta_+^3$ takes finite values at those zeros. Thus, following the rule \eqref{eq:ind_wind}, either $\alpha_+^3$ or $\beta_+^3$ can be used as a local section in the vicinity of these points. In both cases, the index of these zeros is $W_{\bar{k}_4}=+1$, and $W_{\bar{k}_3}=W_{\bar{k}_5}=-1$, yielding consistently a Chern number $C_+=-1$. 
Panels B and C of Fig.~\ref{fig_jauge3} show one additional phase vortex in each individual component of $\alpha_+^3$ and $\beta_+^3$ at different locations in the BZ. However, because the norm $|\psi_+^3(k)|\neq 0$ at those points, they are not considered for the computation of the Chern number.
Similarly, the \textit{tautological bundle method} can be applied directly from panel D showing $\mu_+^3=\alpha_+^3/\beta_+^3$, from which we get $\deg \mu_+^3 = W_{k_2}[\mu_+^3]=+1$ and, thus, $C_+=-1$. Note that we could have equivalently chosen the map $\mu_+^3=\beta_+^3/\alpha_+^3$: compared to panel D, the zero would be at $\bar{k}_1$ but the argument would be flipped, leaving invariant its degree.

\subsection{Remarks concerning normalized eigenvectors}
In quantum crystals, the normalization of the electron wavefunction in real space is $\int dr |\psi^{(n)}_k(r)|^2=1$, which is imposed by the probability of $1$ to find an electron (of a given orbital) in each unit cell, yields the normalization constraint  $\braket{\Psi^{(n)}_k|\psi^{(n)}_k}=|\psi^{(n)}_k|^2=1$ in reciprocal space for every $k$ in BZ, with the standard norm in $\mathbb{C}^N$. This normalization prevents $\psi^{(n)}_k$ from vanishing in BZ, and as a consequence, the \textit{0-section method} becomes meaningless.
From a mathematical point of view, the \textit{normalized} Bloch bundle -- i.e. constructed from the normalized eigenstates -- is not a vector bundle but a $U(1)$-bundle, where $U(1)$ refers physically to the gauge freedom that reduces to a multiplication by a phase factor rather than by an arbitrary complex number of non zero modulus. 
However, those two bundles describing respectively normalized or non-normalized eigenstates over BZ, have isomorphic associated principal bundles \cite{Husemoller}, meaning that the Chern numbers inferred with or without normalizing the eigenvectors are the same. These Chern numbers can be evaluated with the \textit{0-section method} in the non-normalized case, by integrating the Berry curvature in the normalized case, or with the \textit{tautological bundle method} in either case.



\section{Photonic lattice experiment} 

\subsection{Experimental setup} 

To illustrate experimentally the above methods, we study a two-band model with nontrivial Chern bands using a time-multiplexed lattice based on two coupled fibre rings of slightly different lengths (45.34 m and 44.63 m). 
The dynamics of light pulses in this system evolves in time following an evolution operator whose bands of eigenmodes can be engineered to have non-zero Chern numbers.
The set-up is sketched in Fig.~\ref{fig:expt_lattice}(A) and described in detail in Ref.~\cite{asapanna_observation_2025} and in the Supplemental Material~\cite{Supplementary}.

A short light pulse (1.4~ns) injected in one of the rings is subject to a cascade of splitting events into the two rings each time it encounters a beamsplitter (VBS).
The light pulses are amplified within each ring using commercial Er amplifiers to compensate for injection and extraction losses.
Each round trip in the rings is a step in time in the evolution of the pulses, while their temporal ordering within each ring at each time step can be assigned to a synthetic spatial position.
In this way, the dynamics of light pulses in the rings can be mapped onto the evolution of particles in the one-dimensional lattice sketched in Fig.~\ref{fig:expt_lattice}(B) with two sublattices $\alpha$ and $\beta$ --corresponding to sites in each ring, and subject to a discrete-step evolution. The dynamics follow these equations~\cite{Regensburger2011, Bisianov2019, Adiyatullin2023}:
\begin{eqnarray} \label{Eq:step}
\alpha_n^{m+1} &=& \left(\cos\theta_m\alpha_{n-1}^m + i\sin\theta_m\beta_{n-1}^m\right) e^{i\varphi_m} 
\nonumber \\
\beta_n^{m+1} &=& i\sin\theta_m\alpha_{n+1}^m + \cos\theta_m\beta_{n+1}^m
\end{eqnarray}
\noindent with $\alpha_n^{m}$ and $\beta_n^{m}$ being the amplitude of the light pulses at each ring at spatial position $n$ and time step $m$. 
The evolution is governed by the splitting angle $\theta_m$ of the variable beamsplitter connecting the rings, which can be controlled electronically.
A phase modulator in the $\alpha$ ring sets a controlled phase $\varphi_m$ with a value alternating between $\varphi$ and $-\varphi$ at odd and even steps.
The modulator acts as a generalised quasimomentum resulting in a parametric dimension ($\varphi \in (-\pi, \pi]$), and the lattice is effectively two-dimensional with a Brillouin zone given by quasi-momentum $k$ (conjugate of the spatial dimension $n$) and the parameter quasi-momentum $\varphi$.

\begin{figure}
    \includegraphics[width=\columnwidth]{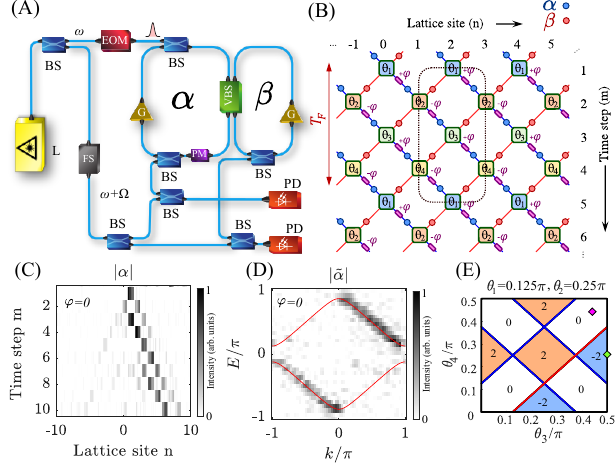}
    \caption{(A) Scheme of the double ring system with injection and reference laser L, beamsplitters BS, variable beamsplitter VBS, frequency shifter FS, phase modulator PM, Erbium doped amplifier G that compensates for losses and photodiode D. 
    (B) Synthetic split-step lattice for the 4-step protocol.
    (C) Measured light intensity in the $\alpha$ ring in an example of step-evolution after injection at a single site for a lattice model with $\theta_1=0.125\pi$, $\theta_2=0.25\pi$, $\theta_3=0.5\pi$ and $\theta_4=0.25\pi$ and $\varphi=0$. 
    (D) Measured band tomography as a function of quasimomenta $k$ and $\varphi$ for the $\theta_i$ parameters of (c). Analytic bands are overlayed in red.
    (E) Bulk topological phases according to the Chern number of the upper band for $\theta_1=0.125\pi$ and $\theta_2=0.25\pi$ as a function of $\theta_3$ and $\theta_4$. Solid lines show the closing of the $E=0$ (blue) and $E=\pi$ (red) gaps.}
    \label{fig:expt_lattice}
\end{figure}

Equation (\ref{Eq:step}) can be written in the form of a series of unitary operators $U_m$ acting on an initial vector state: $\ket{\alpha, \beta}^T_m=U_{m}U_{m-1}\cdots U_1\ket{\alpha, \beta}^T_0$. 
We employ a sequence of four ordered values of the splitting angle $\theta_1$, $\theta_2$, $\theta_3$, $\theta_4$, which is repeated every four steps (i.e., the Floquet period): $U_{F}=U_{4}U_{3}U_{2}U_{1}$.
The Floquet evolution operator in momentum space $\tilde{U}_{F}(k,\varphi)$ is a $2 \times 2$ matrix that has two bands of eigenvalues and eigenvectors characterized by different Chern numbers depending on the values of $\theta_j$, $j=1,\dots,4$ (see Ref.~\cite{asapanna_observation_2025}).
A corner of the topological phase diagram with different Chern phases (computed from the analytical Berry curvature) is displayed in Fig~\ref{fig:expt_lattice}(E).
We can use the mathematical methods described above to classify topologically the bands of eigenvalues of the operator $\tilde{U}_{F}(k,\varphi)$ by analysing their corresponding eigenvectors $\psi_\pm(k,\varphi)=(\tilde{\alpha}_{\pm}(k,\varphi),\tilde{\beta}_{\pm}(k,\varphi))^T$.

\begin{figure*}[t!]
    \includegraphics[width=\textwidth]{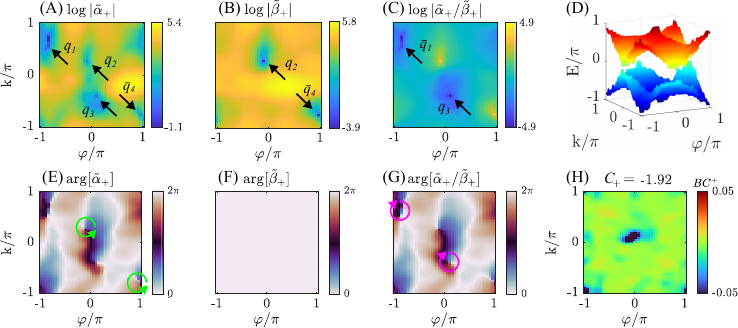}
    \caption{
    \textbf{Upper-band eigenvectors in a topological lattice ($C_+=-2$).}
    Splitting angles are $\theta_1=0.125\pi$, $\theta_2=0.25\pi$, $\theta_3=0.5\pi$, and $\theta_4=0.25\pi$.
    (A,B) Measured (unnormalized) amplitudes and (E,F) corresponding phases of the Floquet--Bloch eigenvectors on the two sublattices, $|\tilde{\alpha}_{+}\rangle$ and $|\tilde{\beta}_{+}\rangle$.
    (C) Amplitude ratio and (G) relative phase between components, $\tilde{\alpha}_{+}/\tilde{\beta}_{+}$.
    Arrows indicate zeros of the unnormalized eigenvectors.
    (D) Measured eigenvalues of the Floquet operator.
    (H) Berry curvature reconstructed from the measured eigenvectors, along with the integrated Chern number of the upper band.
    }
    \label{fig:gauge1_chern2}
\end{figure*}

The eigenvalues and eigenvectors of $\tilde{U}_{F}$ can be experimentally reconstructed from the measurement of the amplitude and phase at each site $n$ and time step $m$ of the light pulses circulating in the two rings.
Figure~\ref{fig:expt_lattice}(C) shows a typical measurement of the spatiotemporal dynamics of light pulses in the the $\alpha$ ring.
By mixing the output signal with a local oscillator (a continuous wave laser detuned by 3~GHz from the input laser wavelength), the eigenvalues and eigenvectors of the lattice dynamics can be extracted by computing the Fourier transform of Fig.~\ref{fig:expt_lattice}(C) for the $\alpha$ ring sites and, analogously, for the $\beta$ ring sites.
The result for the $\alpha$ sublattice is shown in Fig.~\ref{fig:expt_lattice}(D). The complex amplitude of the eigenvector can be directly extracted from the value of the Fourier transform at the coordinates corresponding to the eigenvalue of a given band at a particular value of $k$ and $\varphi$.
By measuring the spatiotemporal dynamics for different values of $\varphi$ in both rings, the complex amplitudes $\tilde\alpha(k,\varphi)$ and $\tilde\beta(k,\varphi)$ of the eigenvectors can be extracted over the entire BZ.
For the analysis presented here, the eigenvectors are reconstructed from the experimental data after Gaussian smoothing, which reduces the noise arising from experimental uncertainties.
The detailed method is described in Refs.~\cite{el_sokhen_edge-dependent_2024, asapanna_observation_2025} and in the Supplemental Material~\cite{Supplementary}.


Panels (D) of Figs.~\ref{fig:gauge1_chern2} and \ref{fig:gauge1_chern0}  show the measured band structures for two different Floquet sequences of splitting angles $\theta_j$, $j="1\dots4"$. Those 2D band structures are gapped so that their Chern numbers are well defined. Panels (A) and (B) display the measured amplitude of the $\tilde{\alpha}_+(k,\varphi)$ and $\tilde{\beta}_+(k,\varphi)$ components of the upper-band eigenvectors in logarithmic scale.
Panels (E) and (F) show the corresponding values of the phase. 
The Berry curvature is then reconstructed from  those measured  amplitudes  and phases of the eigenvectors, and shown in panels (H).  Approximated values of the Chern number are then inferred from the numerical sum of the Berry curvature over the entire BZ, yielding $C_+=-1.92$ and $C_+ = 0.02$ in Figs.~\ref{fig:gauge1_chern2} and \ref{fig:gauge1_chern0}, respectively (see Supplemental Material~\cite{Supplementary}).  
We now apply the two methods discussed above to read out the Chern number without resorting to the Berry curvature.

\subsection{0-section method : experiment}

To extract the value of the Chern number in the upper band using the \textit{0-section method}, we first need to identify the common zeros of the eigenvector components $\tilde{\alpha}_+$ and $\tilde{\beta}_+$ in the Brillouin zone. 
Finding exact zeros from the amplitude of the field alone is not straightforward due to the presence of background experimental noise and electronic noise in the detectors. 
Here, distinct local minima can be clearly identified within the Brillouin zone. 
We focus first on Fig.~\ref{fig:gauge1_chern2}. 
For this set of angle parameters, four minima are observed at $\bar{q}_1=(0.71\pi,-0.85\pi)$, $\bar{q}_2=(0.27\pi,-0.1\pi)$, $\bar{q}_3=(-0.39\pi,0.1\pi)$, and $\bar{q}_4=(-0.75\pi,0.95\pi)$ for $\tilde{\alpha}_+$ and $\bar{q}_2$, $\bar{q}_4$ for $\tilde{\beta}_+$, as indicated by the arrows in panels (A)–(C). 
Among these points, we concentrate on the relevant zeros at $\bar{q}_2$ and $\bar{q}_4$, where both components simultaneously approach zero.
The identification of those minima as actual zeros of the component $\tilde{\alpha}_+$ is made possible as their location coincide with a phase vortex in its argument. The situation is different for $\tilde{\beta}_+$, that is a real-valued function due to the choice of gauge in which the data are analyzed. 
Its argument is thus constant and equal to zero. 
As a consequence, it is not clear whether its two minima are actual zeros. Since their depth is comparable to those of $\tilde{\alpha}_+$, let us therefore assume for the moment that they do correspond to zeros of $\tilde{\beta}_+$ as well.

Next, following \eqref{eq:ind_wind}, we identify the component of the eigenvectors that goes to zero the slowest at those two points 
by plotting  $|\tilde{\alpha}_+/\tilde{\beta}_+|$ (Fig.~\ref{fig:gauge1_chern2}(C)).
If we look at $\bar{q}_2$ and $\bar{q}_4$, we see in Fig.~\ref{fig:gauge1_chern2}(C) that $|\tilde{\alpha}_+/\tilde{\beta}_+|$ tends to a maximum (it should have reached infinity if we had been able to measure absolute zero amplitudes in the experiment). 
Thus, the component that reaches zero the slowest is $\tilde{\alpha}_+$.
The argument of $\tilde{\alpha}_+$ presents a phase vortex at each of those two points with a winding of $-1$, thus contributing to non-zero indices $\ind_{\bar{q}_2,\bar{q}_4}\sigma_+=W_{\bar{q}_2,\bar{q}_4}[\tilde{\alpha}_+]=-1$, that sum up to  $C_+=\sum_j \ind_{\bar{q}_j}\sigma_+=-2$ according to  Eq.~\eqref{eq:chern_ind}.
Note that this lattice model and choice of gauge for the representation of the eigenvectors is an example in which the maximum of the Berry curvature, located at the center and corners of the BZ in Fig.~\ref{fig:gauge1_chern2}(H), does not coincide with the position of the phase vortices in any of the components.

\begin{figure*}[t!]
    \includegraphics[width=\textwidth]{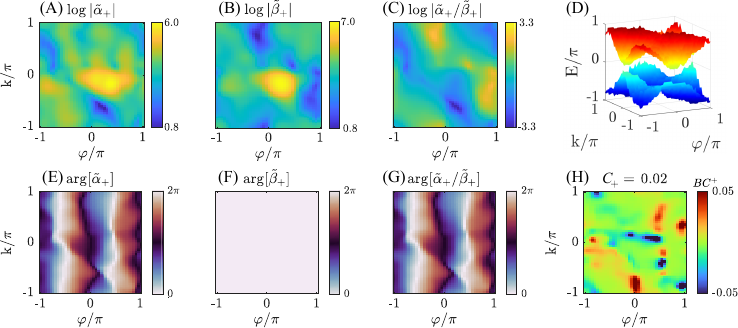}
    \caption{
    \textbf{Upper-band eigenvectors in a topologically trivial lattice ($C=0$).}
    Splitting angles are $\theta_1=0.125\pi$, $\theta_2=0.25\pi$, $\theta_3=0.44\pi$, and $\theta_4=0.44\pi$.
    (A,B) Measured (unnormalized) amplitudes and (E,F) corresponding phases of the Floquet--Bloch eigenvectors on the two sublattices, $|\tilde{\alpha}_{+}\rangle$ and $|\tilde{\beta}_{+}\rangle$.
    (C) Amplitude ratio and (G) relative phase between components, $\tilde{\alpha}_{+}/\tilde{\beta}_{+}$.
    (D) Measured eigenvalues of the Floquet operator.
    (H) Berry curvature reconstructed from the measured eigenvectors, along with the integrated Chern number of the upper band.
    }
    \label{fig:gauge1_chern0}
\end{figure*}

Figure~\ref{fig:gauge1_chern0} presents the same analysis as in Fig.~\ref{fig:gauge1_chern2} for a band model with Chern number equal to zero. 
The amplitude of each component in panels (A) and (B) display a smooth with some minima.
However, $\arg[\tilde{\alpha}_+]$ and $\arg[\tilde{\beta}_+]$ do not show any vorticity in Fig.~\ref{fig:gauge1_chern0}(E) and (F).
From this observation we can already ensure that the winding of any of the components of the eigenvectors at any point in the Brillouin zone must be zero and, necessarily, whichever zero of the eignemodes we could identify would not contribute to the computation of the Chern number using the above method.
We conclude that in this case $C_+ = 0$. 
Representation of the measured eigenvalues in a different gauge in which the eigenvector components exhibit phase vortices, and therefore isolated zeros, even for the topologically trivial phase with $C_+=0$, is presented in the Supplementary Material~\cite{Supplementary}.

These two values of the Chern numbers match the closest integer value obtained from the integration over the BZ of the experimental Berry curvature. 
They are furthermore in agreement with the analytical calculation provided in~\cite{Supplementary}. 

\subsection{Tautological bundle method : experiment}

Let us now extract the Chern number from the experimental data following the \textit{tautological bundle method}.
To do so, we need to identify the zeros of the map $\mu_+=\tilde{\alpha}_+/\tilde{\beta}_+$, whose amplitude is displayed in Fig.~\ref{fig:gauge1_chern2}(C) for the model with $C_+=-2$.
Two minima are found at $\bar{q}_1=(0.71\pi,-0.85\pi)$ and $\bar{q}_3=(-0.39\pi,0.1\pi)$. 
The phase of $\mu_+$, i.e., $\arg[\tilde{\alpha}_+/\tilde{\beta}_+]$, is shown in Fig.~\ref{fig:gauge1_chern2}(G), and it exhibits a winding $W_{\bar{q}_1, \bar{q}_3}[\mu_+]=+1$ around each point, shown by clockwise arrows (pink).
The presence of these phase vortices at points $\bar{q}_1$ and $\bar{q}_3$ justifies the identification of those local minima in the amplitude of $\mu_+$ as zeros of $\mu_+$.
Note that the presence of phase singularities does not alone indicate the presence of zeros, as $\mu_+$ could also be infinite at those points. It follows that the degree of $\mu_+$ is $\deg \mu_+ = \sum_j W_{\bar{q}_j}[\mu_+] = +2$ yielding the Chern number $C_+ = -2$, according to Eq.~\eqref{eq:chern_tauto}.

The amplitude and phase of the map $\mu_+$ is also shown in Fig.~\ref{fig:gauge1_chern0} for the other set of splitting angle parameters. There, the extrema in the amplitude of $\mu_+$ do not come with a phase singularity, implying the non-existence of zeros. The chern number of this band is thus $C_+=0$.  Those two values are consistent with those obtained with the \textit{0-section method}.

\section{Conclusion}

We have clarified the connection between the Chern numbers of the energy  bands of a two-dimensional crystal and the singularities of the Bloch eigenvectors, by providing two mathematically justified and experimentally applicable methods to extract those topological indices from the eigenstates. 
By doing so, we have changed the focus of the analysis of topological properties in wave topology from the Berry curvature, which is extended over the entire BZ and difficult to measure, to a few singularities of the phase of the eigenvector components.
Reference~\cite{fosel_l_2017} discusses a related situation in the context of polarization singularities.
The presence of a phase winding in the Brillouin zone of  eigenvector components (\textit{0-section method}) and of its ratio (\textit{tautological bundle method}) is a necessary condition to have a non-zero Chern number. Given the robustness of the quantized vortical phase distributions at this points, the measurement of the Chern number from their winding is much more precise than the evaluation of the Chern number from the experimental Berry curvature, which implies gradients in amplitude and phase that are significantly more sensitive to noise.

Importantly, associated to these phase singularities, we expect to find absolute zeros of the eigenmodes.
In an experimental context, this has a relevant implication: at these points in the Brillouin zone, it is intrinsically impossible to create a wavepacket with an external stimulation.
Therefore, the topology of the Bloch bands does not only affect the wavepacket dynamics through the appearance of the well-known anomalous velocity at points where the Berry curvature is finite, it also prevents the excitation of certain points of the  Brillouin zone associated to the singularities we have discussed in this work.

Finally, let us stress that the presence of phase singularities is a necessary condition to have a non-zero Chern number, but it is not a sufficient one. In particular, the windings of either the arguments of the eigenvector components (\textit{0-section method}), or of the maps $\mu_\pm$ (\textit{tautological bundle method}), necessarily sum up to zero in the Brillouin zone, due to its compactness. The two methods presented here provide a rigorous and practicable path to identify which phase singularities contribute to the Chern number. And even when those contributions are established, they may in the end still sum up to zero, as we saw for the Haldane model in one of its trivial phases.

\begin{acknowledgments}
This work was supported by the European Research Council grant EmergenTopo (865151), the French government through the Programme Investissement d'Avenir (I-SITE ULNE /ANR-16-IDEX-0004 ULNE) managed by the Agence Nationale de la Recherche, the Labex CEMPI (ANR-11-LABX-0007), and the region Hauts-de-France. 
It was partially funded by the CDP C2EMPI, as well as the French State under the France-2030 program, the University of Lille, the Initiative of Excellence of the University of Lille, the European Metropolis of Lille for their funding and support of the R-CDP-24-004-C2EMPI project.
This project has received funding from the European Union’s Horizon 2020 research and innovation program under the Marie Skłodowska-Curie grant agreement No 101108433.
\end{acknowledgments}

\bibliographystyle{unsrt}
\bibliography{sample}


\begin{appendices}

\widetext

\setcounter{equation}{0}
\setcounter{figure}{0}
\setcounter{table}{0}
\setcounter{page}{1}
\makeatletter
\renewcommand{\theequation}{S\arabic{equation}}
\renewcommand{\thefigure}{S\arabic{figure}}

\section*{Supplementary Material} 
\addcontentsline{toc}{section}{Supplementary Material} 

\section{Experimental set-up}
The experimental setup employs a continuous wave (CW) single-frequency laser source (Koheras MIKRO, NKT Photonics) operating at 1550 nm with a maximum output power of 40 mW and a linewidth $<$ 0.1 kHz. 
The laser output is split equally using a 50/50 beam splitter, with one part serving as input to a local oscillator and the other being modulated into 1.4 ns pulses via an electro-optical modulator (EOM, iXblue MXER-LN-10), which is controlled by an arbitrary waveform generator (AWG 7000B, Tektronix). 
To reduce residual laser light entering the ring, an acoustic optical modulator (AOM, AA Opto-electronic MT110-IIR30-Fio-PM0.5) with an extinction power of -70 dB is incorporated. 
The AOM is shaped in a gate centred in time at the pulse generated by the preceding EOM. 
The prepared injection signal is then introduced into the long $\alpha$ ring through a 70/30 beamsplitter.

The pulse evolution follows a split step walk. 
The two fiber rings $\alpha$ and $\beta$ are coupled via a high-bandwidth 40 GHz electronically controlled variable beamsplitter (EOSpace AX-2x2-0MSS-20). 
Each ring has an Erbium-doped fiber amplifier (EDFA, Keopsys CEFA-C-HG) and an optical variable attenuator (VOA, Agiltron) which are used to finely compensate for round trip losses.

A 90/10 beamsplitter within each ring extracts light for measurement. 
To access both amplitude and phase information of sublattices $\alpha_n^m$ and $\beta_n^m$, a heterodyne measurement technique is employed. 
This involves beating the pulse extracted from the double rings with a local oscillator reference wave.
This wave is derived from the laser used to inject the initial pulse and is frequency-shifted by 3 GHz using an electro-optic modulator.
The beating interference between the signal and the local oscillator is converted to electrical signals using a fast photodiode (Thorlabs DET08CFC, bandwidht 5~GHz). 
These signals are then captured and analyzed using a high-performance oscilloscope (Tektronix MSO64) featuring a 6 GHz bandwidth, 10-bit vertical resolution, 25 GS/s sampling rate, and a memory length of 62.5 Mpts corresponding to 2.5 ms, enabling very detailed signal analysis of the beating.

\begin{figure*}[b!]
    \centering
    \includegraphics[width=\textwidth]{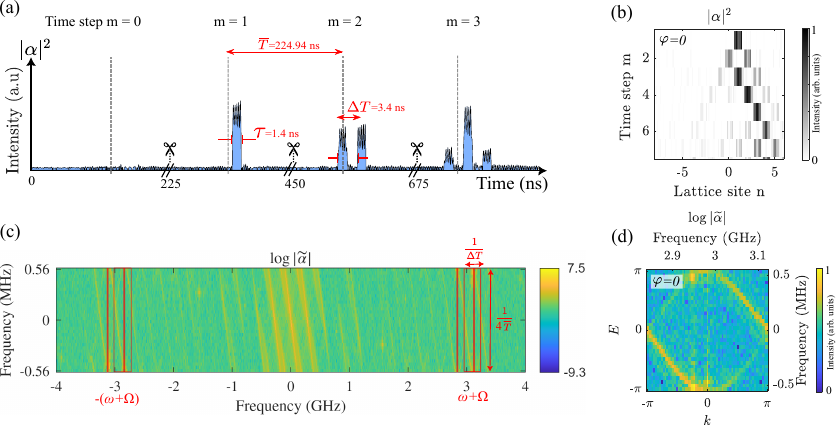}
    \caption{\textbf{Calibration shot.}
    (a) Zoom on the first time steps of the measured time trace of the signal intensity at the output of the $\alpha$ fiber loop. 
    (b) Corresponding spatio-temporal diagram reconstructed from (a), shown here for the initial steps (the full dataset spans over 80 steps).
    (c) Two-dimensional Fourier transform (2DFT) of the stroboscopic spatio-temporal diagram of the $\tilde{\alpha}$.
    (d) Measured bands in one Brillouin zone for the $\tilde{\alpha}$ after zooming the region at $\omega + \Omega$ frequencies in panel (c) after calibration.
    }
    \label{supp_fig:ext_ST_bands}
\end{figure*}

\subsection{Experimental raw eigenvector extraction}

To extract the individual amplitudes and phases of each sublattice, $\alpha$ and $\beta$, the relative amplitude $\alpha/\beta$, and the phase difference $\arg[\alpha/\beta]$ in the experiment, we follow the methodology outlined in Refs.~\cite{asapanna_observation_2025, el_sokhen_edge-dependent_2024}.
Time multiplexed data is collected by recording the output intensity from both fiber loops using photodiodes connected to an oscilloscope. The resulting signal displays groups of pulses separated by the average round-trip time of $\Bar{T}$ = 224.94 ns, with pulses within each group spaced $\Delta T$ = 3.4 ns apart due to the length difference between the two fiber loops as shown in  Fig.~\ref{supp_fig:ext_ST_bands}(a). 
This time trace is then segmented and arranged into a spatio-temporal diagram (Fig.~\ref{supp_fig:ext_ST_bands}(b)) for further analysis. 

The phase information is retrieved by subjecting the signal to optical heterodyne measurement with a reference continuous wave laser, frequency-shifted by about 3 GHz, producing observable fringes in the recorded signal.
The interference between the local oscillator and the signal evolving in the rings contains phase information relevant to the measurement of the band structure (see  the beating signal on top of each pulse in Fig.~\ref{supp_fig:ext_ST_bands}(b)). 
The band structure is reconstructed by performing a numerical two-dimensional Fourier transform (2DFT) on the stroboscopic spatio-temporal diagram of Fig.~\ref{supp_fig:ext_ST_bands}(b) of each ring at time steps corresponding to integer Floquet periods ($m = 4,8,12,\cdots$). 
This yields periodic eigenvalue bands spanning about 10 GHz in the quasimomentum direction this value is fixed by the time resolution of the oscilloscope that records the time trace) and 1.12 MHz in the quasienergy direction, see Fig.~\ref{supp_fig:ext_ST_bands}(c).
We focus on a single Brillouin zone at around a frequency of 3GHz as shown in Fig.~\ref{supp_fig:ext_ST_bands}(d).
The vertical and horizontal axis of the dispersion are then relabelled to span the full spectral Brilluoin zone both in quasienergy $E$ and quasimomentum $k$, spanning both from $-\pi$ to $\pi$.

\begin{figure*}[b!]
    \centering
    \includegraphics[width=0.85\textwidth]{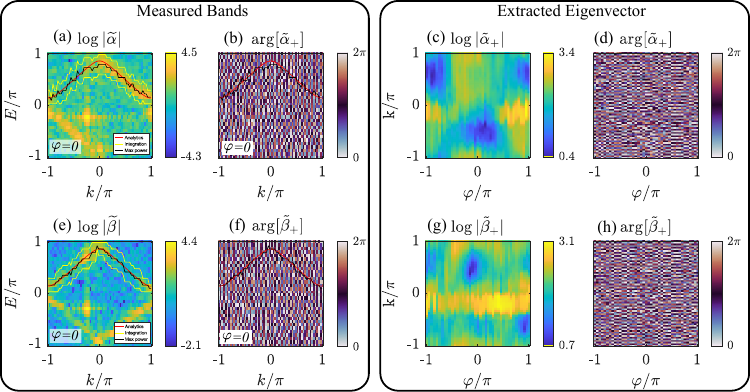}
    \caption{\textbf{Science shot: raw eigenvector extraction $|\psi_{\mathrm{raw}}\rangle$.}
    (a) Measured band structure of $\tilde{\alpha}$ in one Brillouin zone (zoom on the $\omega+\Omega$ region, cf. Fig.~\ref{supp_fig:ext_ST_bands}(d)) at $\varphi=0$ for lattice parameters $\theta_1=0.125\pi$, $\theta_2=0.25\pi$, $\theta_3=0.5\pi$, $\theta_4=0.25\pi$.
    (b) Corresponding phase of $\tilde{\alpha}$.
    (e,f) Band structure and phase of $\tilde{\beta}$.
    (c,d) Amplitude and phase of the extracted upper-band eigenvector component $\tilde{\alpha}_+$ over $\varphi \in (-\pi,\pi]$. 
    (g,h) Same as (c,d) for $\tilde{\beta}_+$.
    }
    \label{supp_fig:ext_EV_raw}
\end{figure*}

Environmental factors can cause fluctuations in fiber length, resulting in shifts of the band structure. 
To diminish these fluctuations we use piezos to lock the lengths of the rings 20~ms before the expriment.
This time is short enough to prevent changes in the length of the rings due to thermal fluctuations.
Even after this compensation we still have slight shifts in band structure due to minor jitters in experiment.
To compensate for these minor shifts, the experimental setup employs a dual-pulse technique to reconstruct and calibrate the two-dimensional band structure of the synthetic lattice. 
This method utilizes two consecutive $\tau$ = 1.4 ns pulses: a calibration shot and a science shot. 
The calibration shot, enters the ring and evolves in time with a constant splitting of 50/50 in the variable beamsplitter and no phase modulation.
The spatio-temporal evolution dynamics produces a well-known reference band structure. 
In contrast, the science shot implements the experimental system of interest, featuring controlled values of the variable beamsplitter angle and phase modulation of phase modulator.
The calibration shot's band structure is compared to its theoretical model, allowing for the measurement of horizontal and vertical shifts Fig.~\ref{supp_fig:ext_ST_bands}(d). 
These measurements are then used to calibrate the axes and offsets, which remain valid for the subsequent science shot measurement. 

For the eigenvector extraction, the quasienergy for each quasimomentum is identified by scanning around the analytically computed bands to locate intensity maxima Fig.~\ref{supp_fig:ext_EV_raw}(a,b) for $\tilde{\alpha}$ and Fig.~\ref{supp_fig:ext_EV_raw}(e,f) for $\tilde{\beta}$ (shown only for upper-band). 
To reduce noise in the measurement of the amplitude, the recorded intensity is integrated over a small range surrounding the band maximum for each $ k $ value as shown in Fig. ~\ref{supp_fig:ext_ST_bands}(a,e). 
Next, the complex amplitudes $\tilde{\alpha}$ and $\tilde{\beta}$ of the measured bands are obtained at each quasimomentum $ k $ for both the $\alpha$ and $\beta$ rings.

This yeilds the raw complex amplitudes of measured eigenvectors at specific quasimomentum $\varphi$.

The calibration shot is crucial for establishing a consistent phase reference across different measurements; specifically, in the calibration shot, the phase is rigidly shifted to zero at $ k = -\pi $, and the same shift phase is applied to the science shot.
The sublattice phase pattern is then reconstructed from independent measurements taken at various values of $\varphi \in (-\pi, \pi]$, utilizing the calibration shot as a reference. 
In this way we obtain the unnormalised raw eigenvectors for the two bands in quasimomentum space:
\begin{equation}\label{eigenvectorsExp}
    {\psi^\mathrm{raw}_{\pm}} =\begin{pmatrix}
        \tilde{\alpha}_{\pm}(k,\varphi) \\ \tilde{\beta}_{\pm}(k,\varphi)
    \end{pmatrix} 
\end{equation}

\subsection{Eigenvector reconstruction using projector smoothing}

As shown in Fig.~\ref{supp_fig:ext_EV_raw}(d,h), the phases of the experimentally reconstructed eigenvectors contain strong point-to-point fluctuations. 
Their origin is multiple. It comes from cumulated noise along the propagation of the wavepacket and inhomogeneities in the amplification compensation, which weakly oscillates at different time steps. In addition, the extracted phase is extremely sensitive to the precise point chosen in the measured band structure [Fig.~\ref{supp_fig:ext_EV_raw}(b,f)] and on the underlying discretisation of the data.
Nevertheless, the two components $\tilde{\alpha}$ and $\tilde{\beta}$ are measured in the same experimental realization and follow the same data analysis protocol and, therefore, acquire approximately the same random contribution to the phase.
To illustrate this idea we can express the measured eigenvector at each point $(k,\varphi)$ as: 
\begin{equation}
{\psi^{\mathrm{raw}}(k,\varphi)}
=
e^{i\chi(k,\varphi)}
\begin{pmatrix}
\tilde{\alpha}(k,\varphi)\\
\tilde{\beta}(k,\varphi)
\end{pmatrix},
\end{equation}
where $\chi(k,\varphi)$ represents the random common phase introduced by the experiment.

Directly smoothing $\tilde{\alpha}$ and $\tilde{\beta}$ is not suitable for two reasons. First, the random phase $\chi(k,\varphi)$ can vary between neighboring data points, even when the underlying physical state varies smoothly. Averaging the complex components would therefore mix eigenvectors carrying different phase factors. Second, the phase of a component becomes undefined when that component vanishes. Standard smoothing of amplitudes and phases can consequently fail near such phase singularities, as discussed in Ref~\cite{guillot_sublattice_2025}.

To avoid these problems, remove the experimental random contribution to the phase, and finally get clean amplitude-phase eigenvectors, we first map each measured eigenvector to the unnormalised rank-one matrix
\begin{equation}
P(k,\varphi)
=
\ket{\psi^{\mathrm{raw}}(k,\varphi)}
\bra{\psi^{\mathrm{raw}}(k,\varphi)}.
\end{equation}
For our two-component eigenvector, this gives
\begin{equation}
P=
\begin{pmatrix}
|\tilde{\alpha}|^2 &
\tilde{\alpha}\tilde{\beta}^{*}
\\
\tilde{\beta}\tilde{\alpha}^{*} &
|\tilde{\beta}|^2
\end{pmatrix}.
\label{eq:projector1}
\end{equation}
This is the \textit{projector operator} built directly from the eigenvectors. For each $(k,\varphi)$, its largest eigenvalue has an associated eigenvector identical to the original eigenvector from which it was constructed.
The main advantage of this representation is that the common experimental random phase cancels. For example,
\begin{equation}
\left(e^{i\chi}\tilde{\alpha}\right)
\left(e^{i\chi}\tilde{\beta}\right)^{*}
=
\tilde{\alpha}\tilde{\beta}^{*}.
\end{equation}
The same cancellation occurs in all four matrix elements. Therefore, $P$ is insensitive to the random global phase shared by the two measured components.

The projector representation is also well defined at zeros of either eigenvector component because at those points the amplitude of that component also vanishes.

Mapping the experimental eigenvectors to $P$ therefore allows the data to be smoothed even near singular points.
For an exact eigenstate, the corresponding rank-one projector onto the band commutes with both the Floquet operator $\tilde U_F$ and the effective Hamiltonian $\tilde H_{\mathrm{eff}}=-i\log \tilde U_F$, and thus contains the complete eigenvector information while discarding the quasienergy eigenvalue.

Once we have constructed the $P$ matrix, each matrix element is smoothed using the same Gaussian kernel in the $k$ and $\varphi$ directions. To respect the periodicity of the parameter space, the data were periodically extended before filtering and then cropped back to the original domain. This gives the smoothed matrix
\begin{equation}
\bar P(k,\varphi)
=
\begin{pmatrix}
\bar P_{11} & \bar P_{12}\\
\bar P_{21} & \bar P_{22}
\end{pmatrix}.
\end{equation}

Because the same real Gaussian kernel is applied to all matrix elements, $\bar P$ is a weighted local average of Hermitian positive-semidefinite matrices. It therefore remains Hermitian and positive semidefinite. The filtering removes random experimental fluctuations while retaining the slowly varying structure of the measured eigenstate.

\begin{figure*}[b!]
    \centering
    \includegraphics[width=0.85\textwidth]
    {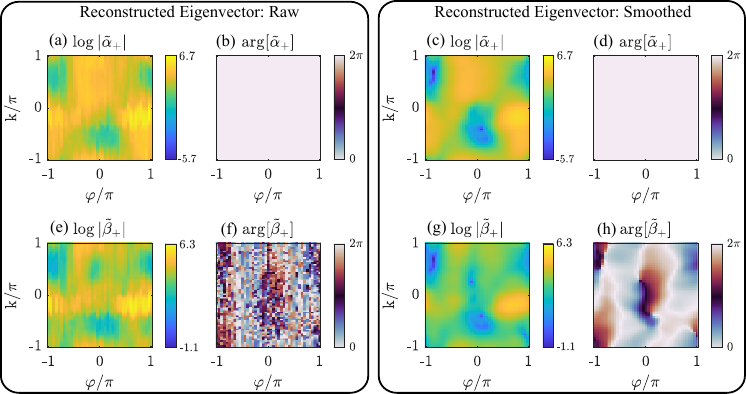}
    \caption{\textbf{Experimental eigenvectors reconstructed in gauge 2 representation, $\psi^{(2)}_{+}$.}
    (a,b) Amplitude and phase of $\tilde{\alpha}_{+}$ reconstructed from the raw projector matrix.
    (e,f) Amplitude and phase of $\tilde{\beta}_{+}$ reconstructed from the raw projector matrix.
    (c,d) and (g,h) Corresponding amplitudes and phases after Gaussian smoothing of the projector elements, using a window of 14 pixels corresponding to $\sigma\approx2.8$ pixels, followed by reconstruction of the dominant eigenstate.}
    \label{supp_fig:ext_EV_gauge2}
\end{figure*}

After smoothing, the (unnormalised) eigenvector is reconstructed analytically from $\bar P$, $\bar P(k,\varphi)=\boldsymbol{a}(k,\varphi)\cdot \boldsymbol{\sigma}$, similar to the main paper. We write
\begin{equation}
\bar P
=
a_0 I
+
a_x\sigma_x
+
a_y\sigma_y
+
a_z\sigma_z,
\end{equation}
where
\begin{align}
a_0 &= \frac{\bar P_{11}+\bar P_{22}}{2},\\
a_x &= \frac{\bar P_{12}+\bar P_{21}}{2},\\
a_y &= \frac{\bar P_{21}-\bar P_{12}}{2i},\\
a_z &= \frac{\bar P_{11}-\bar P_{22}}{2}.
\end{align}
Defining
\begin{equation}
E_{\pm}=\pm\sqrt{a_x^2+a_y^2+a_z^2},
\end{equation}
the eigenvector associated with the largest eigenvalue $a_0+E_{+}$ can be written in either of the following forms, which correspond to the two first gauges discussed in the main text, Eqs.~(5) and (6), respectively:
\begin{equation}
\label{eigenvectorsExp_g1}
{\psi^{(1)}_{+}}
=
\begin{pmatrix}
\tilde{\alpha}_{+}\\
\tilde{\beta}_{+}
\end{pmatrix}
=
\begin{pmatrix}
a_x-i a_y\\
E_{+}-a_z
\end{pmatrix},
\end{equation}
and
\begin{equation}
\label{eigenvectorsExp_g2}
{\psi^{(2)}_{+}}
=
\begin{pmatrix}
\tilde{\alpha}_{+}\\
\tilde{\beta}_{+}
\end{pmatrix}
=
\begin{pmatrix}
E_{+}+a_z\\
a_x+i a_y
\end{pmatrix}.
\end{equation}


Figure~\ref{supp_fig:ext_EV_gauge2} shows the result for the upper band in gauge $2$, Eq.~\eqref{eigenvectorsExp_g2}. The amplitudes and phases reconstructed from the raw projector are shown together with the results obtained after Gaussian smoothing. The procedure strongly reduces the phase fluctuations caused by experimental noise. At the same time, it preserves the phase vortices associated with genuine zeros of the eigenvector components.

The unnormalised eigenvectors reconstructed in gauge $1$, Eq.~\eqref{eigenvectorsExp_g1} are shown in Figs.~4 and 5 of the main text.

\subsection{Analytical discrete-step Floquet operators and Eigenvectors}

The discrete-step walk generated by the experimental setup is described by the coupled equations. 

\begin{eqnarray} \label{coupledEqn}
\alpha_n^{m+1} &=& \left(\cos\theta_m\alpha_{n-1}^m + i\sin\theta_m\beta_{n-1}^m\right) e^{i\varphi_m} 
\nonumber \\
\beta_n^{m+1} &=& i\sin\theta_m\alpha_{n+1}^m + \cos\theta_m\beta_{n+1}^m,
\end{eqnarray}
where $\alpha_{n}^{m}$ and $\beta_{n}^{m}$ are the complex amplitudes of light pulses in the long and short rings, respectively, at time-step $m$ subject to a splitting eveng with angle $\theta$. 
Pulses in the ring $\alpha$ acquire an additional phase due to the phase modulator $\varphi$. 
The system can be understood as a one dimensional quantum walk with a parametric dimension $\varphi$.

In the four-step model, to describe the light dynamics after four time steps, we replace $m$ with $m+3$ in Equation \ref{coupledEqn}. 
We then iterate Equation \ref{coupledEqn} to relate the complex amplitudes of $\alpha$ and $\beta$ at time step $m$ to those at time step $m+3$. 
The coupling angle $\theta$ alternating between four distinct values within a single Floquet period $T_F$ such that $\theta_i$ with $i = m$(mod 4), and $\varphi_m$ alternates between $+\varphi$ and $-\varphi$ for odd and even time steps $m$.
The four-step model exhibits double periodicity: spatial (every two sites $n$ i.e. sites 0, 2, 4, 6, $\cdots$) and temporal (every four-time steps $m$ i.e. steps 4, 8, 12, $\cdots$) as shown in Fig. 3(b) of the main text.
The final equations of motion describing the time evolution during each full four-step period are given by:
\begin{equation}\label{coupledEqn4stepSimp}
    \begin{aligned}
        \alpha_{n}^{m+4} &= T_1T_2T_3T_4\, \alpha_{n-4}^{m} + R_1T_2T_3T_4i\,  \beta_{n-4}^{m} 
        - (R_2R_3T_1T_4e^{i\varphi} + R_1R_2T_3T_4e^{-i\varphi} + R_3R_4T_1T_2e^{-i\varphi})\,  \alpha_{n-2}^{m} \\
        &\quad- (R_1R_2R_3T_4ie^{i\varphi} + R_1R_3R_4T_2ie^{-i\varphi} - R_2T_1T_3T_4ie^{-i\varphi})\,  \beta_{n-2}^{m}
        - (R_1R_3T_2T_4 + R_2R_4T_1T_3 - R_1R_2R_3R_4e^{-2i\varphi})\,  \alpha_{n}^{m} \\
        &\quad- (R_1R_2R_4T_3i - R_3T_1T_2T_4i + R_2R_3R_4T_1ie^{-2i\varphi})\,  \beta_{n}^{m}
        - R_1R_4T_2T_3e^{-i\varphi} \alpha_{n+2}^{m} + R_4T_1T_2T_3ie^{-i\varphi}\,  \beta_{n+2}^{m} \\
        \\
        \beta_{n}^{m+4} &= T_1T_2T_3T_4\,  \beta_{n+4}^{m} + R_1T_2T_3T_4i\,  \alpha_{n+4}^{m}
        \quad- (R_1R_2T_3T_4e^{i\varphi} + R_3R_4T_1T_2e^{i\varphi} + R_2R_3T_1T_4e^{-i\varphi})\,  \beta_{n+2}^{m} \\
        &\quad- (R_1R_3R_4T_2ie^{i\varphi} + R_1R_2R_3T_4ie^{-i\varphi} - R_2T_1T_3T_4ie^{i\varphi})\,  \alpha_{n+2}^{m}
        - (R_1R_3T_2T_4 + R_2R_4T_1T_3 - R_1R_2R_3R_4e^{2i\varphi})\,  \beta_{n}^{m} \\
        &\quad- (R_1R_2R_4T_3i - R_3T_1T_2T_4i + R_2R_3R_4T_1ie^{2i\varphi})\,  \alpha_{n}^{m}
        - R_1R_4T_2T_3e^{i\varphi} \beta_{n-2}^{m} + R_4T_1T_2T_3ie^{i\varphi}\,  \alpha_{n-2}^{m}
    \end{aligned}
\end{equation}
\noindent where $T_{m}=\cos(\theta_{m})$ and $R_{m}=\sin(\theta_{m})$. 

Since the four-step model exhibits double periodicity -- spatial (every two sites $n$) and temporal (every four-time steps $m$)--, we can apply the following Floquet-Bloch ansatz to Eq.~\ref{coupledEqn4stepSimp}:
\begin{equation} \label{floqBloch}
\begin{pmatrix}\alpha_{n}^{m} \\\beta_{n}^{m} \end{pmatrix} 
=\begin{pmatrix}\tilde{\alpha}(k,\varphi) \\ \tilde{\beta}(k,\varphi)  \end{pmatrix} e^{i \frac{E m}{4}} e^{i \frac{k n}{2}}.
\end{equation}
We can rewrite Equation \ref{coupledEqn4stepSimp} as,
\begin{equation}\label{eigenvalueEqn}
    e^{i E} \begin{pmatrix}
        \tilde{\alpha}(k,\varphi) \\ \tilde{\beta}(k,\varphi)
    \end{pmatrix} = \tilde{U}_F \begin{pmatrix}
        \tilde{\alpha}(k,\varphi) \\ \tilde{\beta}(k,\varphi)
    \end{pmatrix},
\end{equation}

\noindent where $\tilde{U}_F$ is the four-step Floquet operator in reciprocal space and Eq.~\ref{eigenvalueEqn} represents the eigenvalue equation for our system with
\begin{equation}\label{UFunitcell}
    \tilde{U}_F(k,\varphi) = \begin{pmatrix}
        \tilde{u}_{1} & \tilde{u}_{2} \\ \tilde{u}_{3} & \tilde{u}_{4}
    \end{pmatrix}
\end{equation}
where, 
\begin{equation}
    \begin{aligned}
        \tilde{u}_{1} &= T_1 T_2 T_3 T_4 e^{-2ik} - R_2 R_4 T_1 T_3 - R_1 R_3 T_2 T_4 + R_1 R_2 R_3 R_4 e^{-2i\varphi} - R_1 R_4 T_2 T_3 e^{ik-i\varphi} - R_2 R_3 T_1 T_4 e^{-ik+i\varphi} \\&\quad- R_1 R_2 T_3 T_4 e^{-ik-i\varphi} - R_3 R_4 T_1 T_2 e^{-ik-i\varphi}\\
        \tilde{u}_{2} &= R_3 T_1 T_2 T_4 i - R_1 R_2 R_4 T_3 i + R_1 T_2 T_3 T_4 i e^{-2ik} - R_2 R_3 R_4 T_1 i e^{-2i\varphi} - R_1 R_2 R_3 T_4 i e^{-ik} e^{i\varphi} - R_1 R_3 R_4 T_2 i e^{-ik} e^{-i\varphi} \\&\quad+ R_4 T_1 T_2 T_3 i e^{ik} e^{-i\varphi} + R_2 T_1 T_3 T_4 i e^{-ik} e^{-i\varphi}\\
        \tilde{u}_{3} &= R_3 T_1 T_2 T_4 i - R_1 R_2 R_4 T_3 i + R_1 T_2 T_3 T_4 i e^{2ik} - R_2 R_3 R_4 T_1 i e^{2i\varphi} - R_1 R_3 R_4 T_2 i e^{ik} e^{i\varphi} - R_1 R_2 R_3 T_4 i e^{ik} e^{-i\varphi} \\&\quad+ R_2 T_1 T_3 T_4 i e^{ik} e^{i\varphi} + R_4 T_1 T_2 T_3 i e^{-ik} e^{i\varphi}\\
        \tilde{u}_{4} &= T_1 T_2 T_3 T_4 e^{2ik} - R_2 R_4 T_1 T_3 - R_1 R_3 T_2 T_4 + R_1 R_2 R_3 R_4 e^{2i\varphi} - R_1 R_2 T_3 T_4 e^{ik} e^{i\varphi} - R_3 R_4 T_1 T_2 e^{ik} e^{i\varphi} - R_1 R_4 T_2 T_3 e^{-ik} e^{i\varphi} \\&\quad- R_2 R_3 T_1 T_4 e^{ik} e^{-i\varphi}
    \end{aligned}
\end{equation}

Solving Equation \ref{eigenvalueEqn} for the eigenvalues of $\tilde{U}_F$ we obtain the solutions of energies of the bands $E_{\pm}$:

\begin{equation}
\begin{aligned}
    E_{\pm} (k,\varphi) &= \pm \cos^{-1}[T_1 T_2 T_3 T_4 \cos(2k) - R_2 R_4 T_1 T_3 - R_1 R_2 T_3 T_4 \cos(k+\varphi) - R_3 R_4 T_1 T_2 \cos(k+\varphi) - R_1 R_4 T_2 T_3 \cos(k-\varphi) \\&\quad- R_2 R_3 T_1 T_4 \cos(k-\varphi) - R_1 R_3 T_2 T_4 + R_1 R_2 R_3 R_4 \cos(2\varphi)],
\end{aligned}
\end{equation}

The (unnormalized) eigenvectors are obtained from the Pauli decomposition of the effective Hamiltonian exactly same as the main paper, 
\begin{equation}
H = -i \log \tilde{U}_F = \vec{h}\cdot\vec{\sigma},
\qquad 
\vec{h} = (h_x,h_y,h_z),
\end{equation}
where \(H\) is Hermitian and can be expressed in terms of Pauli matrices $\vec{\sigma}$.

The eigenvalues are given by
\begin{equation}
E_{\pm} = \pm \|\vec{h}\| \equiv \pm E.
\end{equation}

The corresponding (unnormalized) eigenvectors can be written in two equivalent gauges,
\begin{equation}\label{th_gauge1}
\psi_{\pm}^{(1)} =
\begin{pmatrix} 
h_x - i h_y \\ 
E_{\pm} - h_z 
\end{pmatrix},
\end{equation}

\begin{equation}\label{th_gauge2}
\psi_{\pm}^{(2)} =
\begin{pmatrix} 
E_{\pm} + h_z \\ 
h_x + i h_y 
\end{pmatrix}.
\end{equation}

Another valid gauge is ${\psi_{\pm}^{(3)}}$,
\begin{equation}\label{th_gauge3}
{\psi_{\pm}^{(3)}}={\psi_{\pm}^{(1)}}+{\psi_{\pm}^{(2)}}
\end{equation}

To find the corresponding bulk topological invariant, we first check that $\tilde{U}_F(k,\varphi)$ is in the D symmetry class with particle-hole symmetry
\begin{equation}\label{particleHoleSymm}
    \mathcal{P}\tilde{U}_F(k,\varphi)\mathcal{P}^{-1}=\tilde{U}_F(-k,-\varphi),
\end{equation}
implemented by the anti-unitary operator $\mathcal{P}=\sigma_z K$, being $K$ the complex conjugation operation. This implies that the bulk topological invariant corresponds to the Chern number~\cite{Ryu_2010}.

\subsection{Experimental computation of the Chern number in gauge 3}

The number and positions of the zeros of a section ${\psi^j_\pm}$ depend on the chosen gauge $j$. 
In the main text, the analysis was performed using gauge $1$ (Eq.~\ref{eigenvectorsExp_g1}) of the experimentally reconstructed eigenvectors. 
In that representation, several of the relevant zeros appear close to the Dirac points of the Floquet band structure. 
This should not be interpreted as a general property: the locations of the section zeros are gauge dependent and may be displaced under a smooth gauge transformation, while the total index and hence the Chern number remain unchanged.

\begin{figure*}[t!]
    \centering
    \includegraphics[width=\textwidth]{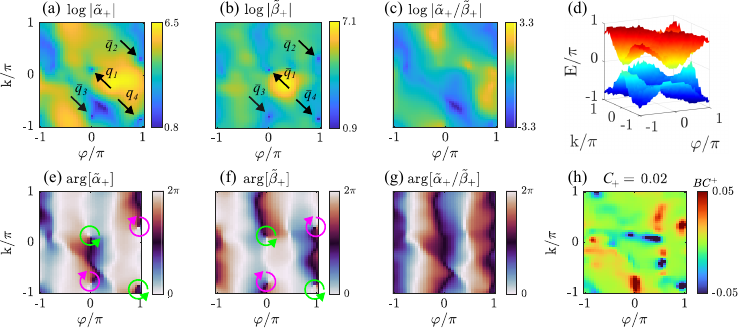}
    \caption{
    \textbf{Experimental upper-band eigenvectors in gauge 3 in a topologically trivial lattice ($C=0$).}
    Splitting angles are $\theta_1=0.125\pi$, $\theta_2=0.25\pi$, $\theta_3=0.44\pi$, and $\theta_4=0.44\pi$.
    (a,b) Measured (unnormalized) amplitudes and (e,f) corresponding phases of the Floquet--Bloch eigenvectors on the two sublattices, $\tilde{\alpha}_{+}$ and $\tilde{\beta}_{+}$.
    (c) Amplitude ratio and (g) relative phase between components, $\tilde{\alpha}_{+}/\tilde{\beta}_{+}$.
    Arrows indicate zeros of the unnormalized eigenvectors; phases at the relevant zeros are used for analysis.
    (d) Measured eigenvalues of the Floquet operator.
    (h) Berry curvature reconstructed from the measured eigenvectors, along with the integrated Chern number of the upper band.
    }
    \label{supp_fig:gauge3_chern0_expt}
\end{figure*}

To illustrate this gauge freedom explicitly, we repeat the analysis of the experimental eigenvectors using the gauge $\psi_\pm^{(3)}=\psi_\pm^{(1)}+\psi_\pm^{(2)}$, for which the relevant zeros of the section do not coincide with the zeros of $\mu$. 
The \emph{0-section method} described in the main text can then be applied in exactly the same manner to determine the Chern number.
Figures~\ref{supp_fig:gauge3_chern0_expt}(a,b) show the measured band structure in that gauge for the Floquet sequence $\theta_1=0.125\pi$, $\theta_2=0.25\pi$, $\theta_3=0.44\pi$, and $\theta_4=0.44\pi$, corresponding to a topologically trivial phase with $C_+=0$. 
Panels (a,b) display the amplitudes, while (e,f) show the corresponding phases of the $\tilde{\alpha}_+$ and $\tilde{\beta}_+$ components of the upper-band eigenvectors. 
Several minima appear in $|\tilde{\alpha}|$, $|\tilde{\beta}|$, and $|\tilde{\alpha}/\tilde{\beta}|$ (indicated by arrows at $\bar{q}=(k,\varphi)$), consistent with zeros within experimental resolution.
We observe four zeros of $\psi$ at $\bar{q}_1=(0\pi,0.1\pi)$, $\bar{q}_2=(0.3\pi,0.95\pi)$, $\bar{q}_3=(-0.8\pi,0\pi)$, and $\bar{q}_4=(-0.84\pi,0.95\pi)$ for both $\tilde{\alpha}_+$ and $\tilde{\beta}_+$ [Figs.~\ref{supp_fig:gauge3_chern0_expt}(a,b)]. 
At these points ($\bar{q}_{1,2,3,4}$), $|\tilde{\alpha}_+/\tilde{\beta}_+|\to \mathbb{C}\setminus\{0\}$.
Therefore, according to relations in Eq.~(4) of the main text, either component can be used for the trivialisation associated to the 0-section method.
Choosing $\alpha$, we obtain $\mathrm{Ind}_{\bar{q}_1,\bar{q}_4}\sigma_+ = W_{\bar{q}_1,\bar{q}_4}[\tilde{\alpha}_+] = -1$  and $\mathrm{Ind}_{\bar{q}_2,\bar{q}_3}\sigma_+ = W_{\bar{q}_2,\bar{q}_3}[\tilde{\alpha}_+] = +1$. 
The corresponding phase map [Fig.~\ref{supp_fig:gauge3_chern0_expt}(f)] shows counter clockwise green arrows for $-1$, and clockwise pink arrows for $+1$. 
Summing these indices via Eq.~(3) yields $$C_+ = \sum_j \mathrm{Ind}_{\bar{q}_j}\sigma_+ = 0$$

\begin{figure*}[t!]
    \centering
    \includegraphics[width=\textwidth]{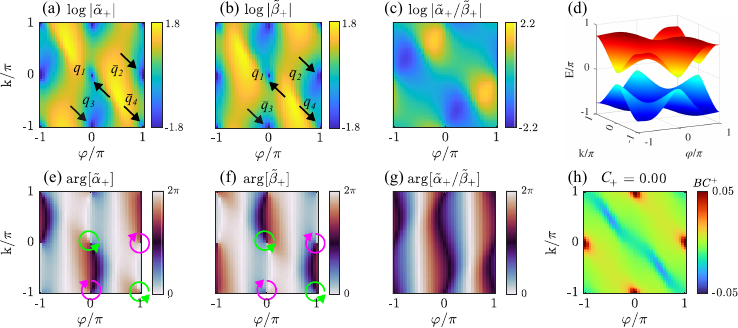}
    \caption{
    \textbf{Analytical upper-band eigenvectors in gauge 3 in a topologically trivial lattice ($C=0$).}
    Splitting angles are $\theta_1=0.125\pi$, $\theta_2=0.25\pi$, $\theta_3=0.44\pi$, and $\theta_4=0.44\pi$.
    (a,b) Theoretical (unnormalized) amplitudes and (e,f) corresponding phases of the Floquet--Bloch eigenvectors on the two sublattices, $\tilde{\alpha}_{+}$ and $\tilde{\beta}_{+}$.
    (c) Amplitude ratio and (g) relative phase between components, $\tilde{\alpha}_{+}/\tilde{\beta}_{+}$.
    Arrows indicate zeros of the unnormalized eigenvectors; phases at the relevant zeros are used for analysis.
    (d) Theoretical eigenvalues of the Floquet operator.
    (h) Berry curvature reconstructed from the theoretical eigenvectors, along with the integrated Chern number of the upper band.
    }
    \label{supp_fig:gauge3_chern0_theory}
\end{figure*}

Analitically computed eigenvectors in the same gauge for the same band are shown in Fig.~\ref{supp_fig:gauge3_chern0_theory}. 
The simulatenaous zeros of both components of $\psi$ are at $\bar{q}_1=(0\pi,0\pi)$, $\bar{q}_2=(0\pi,1\pi)$, $\bar{q}_3=(-1\pi,0\pi)$, and $\bar{q}_4=(-1\pi,1\pi)$ for both $\tilde{\alpha}_+$ and $\tilde{\beta}_+$ [Figs.~\ref{supp_fig:gauge3_chern0_theory}(a,b)]. 
They are slightly shifted from the experimentally measured one due to experimental and the smoothing parameters used for the experimental eigenvecotrs. 
The 0-section method applied to the analytical eigenvectors also produces a total chern number $C_+ = 0$.

\begin{figure*}[t!]
    \centering
    \includegraphics[width=\textwidth]{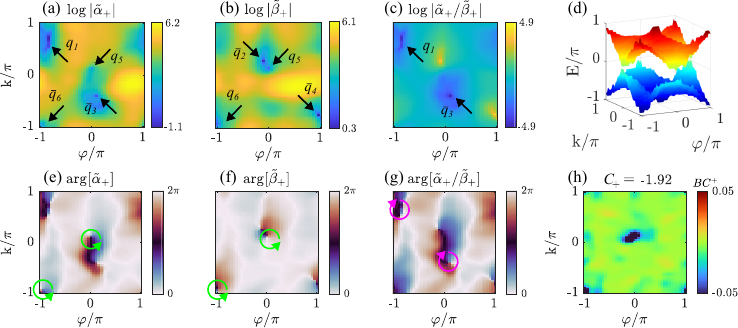}
    \caption{
    \textbf{Experimental upper-band eigenvectors in gauge 3 in a topological lattice with $C_+=-2$.}
    Splitting angles are $\theta_1=0.125\pi$, $\theta_2=0.25\pi$, $\theta_3=0.5\pi$, and $\theta_4=0.25\pi$.
    (a,b) Measured (unnormalized) amplitudes and (e,f) corresponding phases of the Floquet--Bloch eigenvectors on the two sublattices, $\tilde{\alpha}_{+}$ and $\tilde{\beta}_{+}$.
    (c) Amplitude ratio and (g) relative phase between components, $\tilde{\alpha}_{+}/\tilde{\beta}_{+}$.
    Arrows indicate zeros of the unnormalized eigenvectors; phases at the relevant zeros are used for analysis.
    (d) Measured eigenvalues of the Floquet operator.
    (h) Berry curvature reconstructed from the measured eigenvectors, along with the integrated Chern number of the upper band.
    }
    \label{supp_fig:gauge3_chern2_expt}
\end{figure*}

We now turn to a topologically nontrivial phase. 
Figure~\ref{supp_fig:gauge3_chern2_expt} shows the same analysis in gauge 3 for a Floquet sequence with $\theta_1=0.125\pi$, $\theta_2=0.25\pi$, $\theta_3=0.44\pi$, and $\theta_4=0.44\pi$, corresponding to $C_+=-2$. 
In this case, each component of $\psi$ displays a different number of zeros: for $\tilde{\alpha}_+$ there are four zeros at $\bar{q}_1=(0.71\pi,-0.85\pi)$, $\bar{q}_5=(0.0\pi,0.0\pi)$, $\bar{q}_3=(-0.39\pi,0.1\pi)$, and $\bar{q}_6=(-1\pi,-1\pi)$, while for $\tilde{\beta}_+$ there are only two zeros at $\bar{q}_2=(0.26\pi,-0.1\pi)$ and $\bar{q}_4=(-0.74\pi,0.95\pi)$ [Figs.~\ref{supp_fig:gauge3_chern2_expt}(a,b)].  
At the points where both components vanish ($\bar{q}_{5}$ and $\bar{q}_{6}$), we can use either component in Eq.~(4) of the main text for the local trivilisation. 
Choosing $\alpha$, we find $\mathrm{Ind}_{\bar{q}_5,\bar{q}_6}\sigma_+ = W_{\bar{q}_5,\bar{q}_6}[\tilde{\alpha}_+] = -1$. 
The phase map [Fig.~\ref{supp_fig:gauge3_chern2_expt}(e)] shows the winding at these points, leading to $$C_+ = \sum_j \mathrm{Ind}_{\bar{q}_j}\sigma_+ = -2$$

\begin{figure*}[t!]
    \centering
    \includegraphics[width=\textwidth]{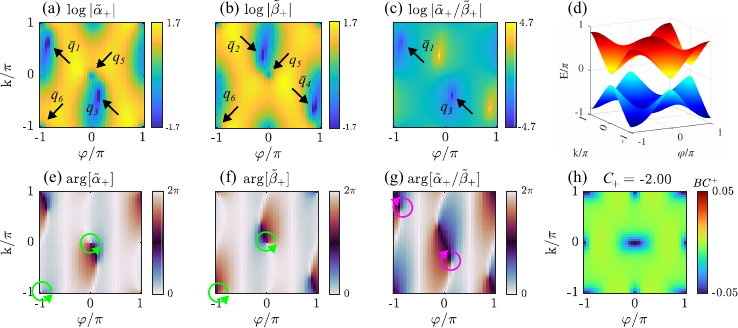}
    \caption{
    \textbf{Analytical upper-band eigenvectors in gauge 3 in a lattice with $C_+=-2$.}
    Splitting angles are $\theta_1=0.125\pi$, $\theta_2=0.25\pi$, $\theta_3=0.5\pi$, and $\theta_4=0.25\pi$.
    (a,b) Theoretical (unnormalized) amplitudes and (e,f) corresponding phases of the Floquet--Bloch eigenvectors on the two sublattices, $\tilde{\alpha}_{+}$ and $\tilde{\beta}_{+}$.
    (c) Amplitude ratio and (g) relative phase between components, $\tilde{\alpha}_{+}/\tilde{\beta}_{+}$.
    Arrows indicate zeros of the unnormalized eigenvectors; phases at the relevant zeros are used for analysis.
    (d) Theoretical eigenvalues of the Floquet operator.
    (h) Berry curvature reconstructed from the theoretical eigenvectors, along with the integrated Chern number of the upper band.
    }
    \label{supp_fig:gauge3_chern2_theory}
\end{figure*}

The corresponding analytical eigenvectors in gauge 3 are displayed in Fig.~\ref{supp_fig:gauge3_chern2_theory}. 
The zeros of the $\tilde{\alpha}_+$ component are located at $\bar{q}_1=(0.68\pi,-0.85\pi)$, $\bar{q}_5=(0.0\pi,0.0\pi)$, $\bar{q}_3=(-0.36\pi,0.12\pi)$, and $\bar{q}_6=(-1\pi,-1\pi)$; for the  $\tilde{\beta}_+$ component at $\bar{q}_2=(0.36\pi,-0.12\pi)$ and $\bar{q}_4=(-0.68\pi,0.85\pi)$ [Figs.~\ref{supp_fig:gauge3_chern2_theory}(a,b)]. 
Using the same trivialisation as in the experiment at $\bar{q}_{5}$ and $\bar{q}_{6}$, we find a total chern number $C_+ = -2$.

\subsection{Computation of Berry curvature and Chern number}

Experimentally, a normalized eigenvector can be constructed from the raw eigenvector of Eq.~\ref{eigenvectorsExp} (or from any gauge-equivalent representation) as

\begin{equation}
\label{eigenvectorsExpNorm}
{\psi_{\mathrm{norm}}}
=
\frac{1}{\sqrt{1+|R(k,\varphi)|^2}}
\begin{pmatrix}
|R(k,\varphi)|e^{i\Phi_{\alpha\beta}(k,\varphi)} \\
1
\end{pmatrix},
\end{equation}

where

\begin{equation}
R(k,\varphi)
=
\frac{\tilde{\alpha}(k,\varphi)}
{\tilde{\beta}(k,\varphi)}
=
|R(k,\varphi)|e^{i\Phi_{\alpha\beta}(k,\varphi)}.
\end{equation}

Figures~4(a,b) and 4(e,f) of the main text display the experimentally measured amplitudes and phases of the $\tilde{\alpha}_{+}$ and $\tilde{\beta}_{+}$ components for the upper band in the topologically trivial phase ($C_{+}=0$). The corresponding results in the gauge defined by Eq.~\ref{th_gauge3} are shown in Fig.~\ref{supp_fig:gauge3_chern0_expt}(a,b) and (e,f). Similarly, the experimentally measured amplitudes and phases for the topological phase with $C_{+}=-2$ are shown in Fig.~5 and in Fig.~\ref{supp_fig:gauge3_chern2_expt} for the alternative gauge.

The Berry curvature is evaluated on the natural discretization of the Brillouin zone provided by the experimental data. For each plaquette, the Berry flux is computed using the gauge-invariant lattice formulation of Fukui, Hatsugai, and Suzuki \cite{tutorial_bc},
\begin{equation}
BC
=
-\mathrm{Im}
\log
\left[
\langle \psi_1|\psi_2\rangle
\langle \psi_2|\psi_3\rangle
\langle \psi_3|\psi_4\rangle
\langle \psi_4|\psi_1\rangle
\right],
\end{equation}
where $|\psi_i\rangle$ denotes the eigenvector at the four corners of the plaquette.

Although Eq.~\ref{eigenvectorsExpNorm} provides a normalized representation of the experimentally reconstructed state, the Berry curvature obtained from the above expression is unchanged if one instead uses the unnormalized experimental eigenvectors. Since the normalisation factor is strictly real and positive, it does not contribute to the phase of the plaquette product. Consequently, the Berry curvature and the resulting Chern number are identical whether one uses normalized eigenvectors or the raw experimental eigenvectors, provided that the eigenvector does not vanish on the plaquette. Normalization is therefore employed only for convenience and numerical stability.

The Chern number is obtained by summing the Berry curvature over the Brillouin zone,

\begin{equation}
C
=
\frac{1}{2\pi}
\sum_{\mathrm{BZ}}
BC .
\end{equation}

Figures~4(h) and 5(h) of the main text show the experimentally measured Chern numbers obtained from the reconstructed eigenvectors. The corresponding results in the alternative gauge are shown in Fig.~\ref{supp_fig:gauge3_chern0_expt}(h) and Fig.~\ref{supp_fig:gauge3_chern2_expt}(h). For comparison, the analytically calculated Chern numbers obtained from the theoretical eigenvectors of Eq.~\ref{th_gauge3} are presented in Fig.~\ref{supp_fig:gauge3_chern0_theory}(h) and Fig.~\ref{supp_fig:gauge3_chern2_theory}(h). The agreement between the different gauges confirms the gauge invariance of the measured topological invariant.

\end{appendices}

\end{document}